# PHOTOMETRY OF SELECTED OUTER MAIN BELT ASTEROIDS

V. G. Shevchenko[a,b*], O. I. Mikhalchenko[a,b], I. N. Belskaya[a,b], I. G. Slyusarev[a,b], V. G. Chiorny[a], Yu. N. Krugly[a], T. A. Hromakina[a], A. N. Dovgopol[a], N. N. Kiselev[c], A. N. Rublevsky[c], K. A. Antonyuk[c], A. O. Novichonok[d], A. V. Kusakin[e], I. V. Reva[e], R. Ya. Inasaridze[f,g], V.V. Ayvazian[f,g], G.V. Kapanadze[f,g], I. E. Molotov[h], D. Oszkiewicz[i], T. Kwiatkowski[i]

[a]Institute of Astronomy of V.N. Karazin Kharkiv National University, Kharkiv 61022, 4 Svobody sq., Ukraine

[b]Department of Astronomy and Space Informatics of V.N. Karazin Kharkiv National University, Kharkiv 61022, 4 Svobody sq., Ukraine

[*]Corresponding Author E-mail address: shevchenko@astron.kharkov.ua

[c]Crimean Astrophysical Observatory, Nauchnij, Crimea

[d]Petrozavodsk State University, Lenin Str. 33, 185910, Petrozavodsk, Republic of Karelia, Russia,

[e]Fesenkov Astrophysical Institute, Observatory 23, 050020 Almaty, Kazakhstan

[f]Kharadze Abastumani Astrophysical Observatory, Ilia State University, K. Cholokoshvili Av. 3/5, Tbilisi 0162, Georgia

[g]Samtskhe-Javakheti State University, Rustaveli Street 113, Akhaltsikhe 0080, Georgia

[h]Keldysh Institute of Applied Mathematics, RAS, 4 Miusskaya sq., Moscow 125047, Russia

[i]Astronomical Observatory Institute, Faculty of Physics, A. Mickiewicz University, Słoneczna 36, 60-286 Poznan, Poland

Pages: 41

Tables: 3

Figures: 23

**ABSTRACT**

We present new photometric observations for twelve asteroids ((122) Gerda, (152) Atala, (260) Huberta, (665) Sabine, (692) Hippodamia, (723) Hammonia, (745) Mauritia, (768) Struveana, (863) Benkoela, (1113) Katja, (1175) Margo, (2057) Rosemary) from the outer part of the main belt aimed to obtain the magnitude-phase curves and to verify geometric albedo and taxonomic class based on their magnitude-phase behaviors. The measured magnitude-phase relations confirm previously determined composition types of (260) Huberta (C-type), (692) Hippodamia (S-type) and (1175) Margo (S-type). Asteroids (665) Sabine and (768) Struveana previously classified as X-type show phase-curve behavior typical for moderate-albedo asteroids and may belong to the M-type. The phase-curve of (723) Hammonia is typical for the S-type which contradicts the previously determined C-type. We confirmed the moderate-albedo of asteroids (122) Gerda and (152) Atala, but their phase-curves are different from typical for the S-type and may indicate more rare compositional types. Based on magnitude-phase behaviors and *V-R* colors, (2057) Rosemary most probably belongs to M-type, while asteroids (745) Mauritia and (1113) Katja belong to S-complex. The phase curve of the A-type asteroid (863) Benkoela does not cover the opposition effect range and further observations are needed to understand typical features of the phase-curves of A-type asteroids in comparison with other types. We have also determined lightcurve amplitudes of the observed asteroids and obtained new or improved values of the rotation periods for most of them.

**Key words**: Asteroids, rotation; Photometry; Spectrophotometry;

## 1. Introduction

It was traditionally believed that the outer part of the main belt asteroids contains mostly dark primitive asteroids with geometric albedo (hereafter, albedo) values up to 0.10. However, the albedos obtained from the infrared surveys (Tedesco et al., 2002; Masiero et al., 2011; Usui et al., 2011), taxonomic distribution of asteroids based on SDSS color indices and albedos (Carvano et al. 2010; DeMeo, Carry, 2013), phase curve slopes from low quality photometric phase curves (Oszkiewicz et al. 2012), and some spectral data (Bus, Binzel 2002; Fornasier et al. 2010; Lazarro et al. 2004; Iwai et al. 2017, etc.) indicate a presence of moderate and high-albedo asteroids in the outer asteroid belt (hereafter we will refer to such objects as high-albedo asteroids). Knowledge of the fraction of such asteroids in this region is very important for understanding the formation and evolution of the asteroid belt and our Solar System as a whole. Kasuga et al. (2013, 2015) performed spectral observations of some high albedo outer-belt asteroids and suggested that the presence of some amorphous Mg-pyroxenes or orthopyroxenes might explain the high albedos of their surfaces. It should be noted that the radiometric albedos might have rather large uncertainties, mainly due to incorrect estimations of the absolute magnitude (Pravec et al. 2012; Shevchenko et al. 2014). As an example, the albedo values of (768) Struveana from two different AKARI data sets differ about two times (0.14, 0.24) due to the use of the same absolute magnitude (Ali-Lagoa et al. 2018). Thus, an independent check of the albedos of such objects should be made. Complementary information on albedo and taxonomic classes of asteroids can be obtained from their magnitude-phase dependences as it was shown in Belskaya, Shevchenko (2000, 2018), Oszkiewicz et al. (2012, 2021), Shevchenko et al. (2015), and Carbognani et al. (2019). The opposition effect in brightness is more pronounced for the high and moderate albedo asteroids while the linear slope of the magnitude-phase relations increases for the low albedo surfaces.

We initiated the observational program to provide an independent verification of albedo and taxonomic class of selected outer-belt asteroids based on their high quality magnitude-phase dependences. In this paper, we present new photometric observations for twelve asteroids that are located in the outer part of the main belt, with the orbital semi-major axis in the range of 2.99-3.39 au. A majority of the observed objects have albedos of 0.2 and higher - according to the IRAS, WISE and AKARI infrared surveys (Tedesco et al., 2002; Masiero et al., 2011, 2012, 2014; Usui et al., 2011; Alí-Lagoa et al., 2018). The diameters of these objects range from 12 to 95 km, with an average diameter of about 45 km.

The results of observations are presented in Section 2. In Section 3, we compare the magnitude-phase relations of the investigated objects with the average values obtained for high, moderate and low albedo asteroids. Using the correlation between the linear coefficient and albedo we have also determined the albedos of our asteroid set. In the Conclusion, we present a short summary of our work.

## 2. Observations and Results

CCD-photometry was carried out in the *BVR* standard bands of Johnson-Cousins photometric system using mainly the 0.7-m reflector of the Institute of Astronomy of V.N. Karazin Kharkiv National University. Some observations were also made at other observational sites with the 0.5-m, 1-m and 1.25-m telescopes of the Crimean Astrophysical Observatory, the 0.7-m telescope of the Kharadze Abastumani Astrophysical Observatory in Georgia, and the 1-m telescope of the Observatory of Fesenkov Astrophysical Institute in Kazakhstan, to minimize influence of weather conditions and obtain a good phase angle coverage. The CCD-image data were reduced with the synthetic aperture photometry package (ASTPHOT) developed at the DLR by S. Mottola (Mottola et al., 1995). The absolute calibrations of the comparison stars were performed with the standard star sequences from Landolt (1992) and Skiff (2007). In some cases for calibration of the comparison stars we used their stellar magnitudes in the SDSS system, taken from the APASS DR9 (Henden et al., 2012) and Pan-STARRS DR1 (Chambers et al., 2016) catalogs. To transform them to the standard *BVR* Johnson-Cousins photometric system we used the corresponding equations given in Fukugita et al. (1996), and Tonry et al. (2012). The accuracy of the obtained absolute photometry is in the range of 0.01-0.03 mag. More details on CCD observations and data reduction methods are given in Krugly et al. (2002), and Shevchenko et al. (2012).

The accurate magnitude-phase relations can be obtained only by taking into account magnitude changes due to the asteroid's rotation. Because of that, we obtained lightcurves for each asteroid, and using our long-term observations, we were able to determine accurate rotation periods for 10 out of 12 of our targets. For each asteroid, our observations are presented as a composite lightcurve constructed according to the procedures described by Harris and Lupishko (1989), and Magnusson and Lagerkvist (1990). The phase angle in parentheses on the magnitude axis indicates the magnitude scale at the initial date of the composite lightcurve. The data for individual nights are denoted by different symbols. They were shifted along the magnitude axis in order to obtain the best fit of a lightcurve. The values of the magnitude shifts for each date, the rotation period as well as the time of the zero phase used to calculate rotational phases are also given.

The magnitude-phase curves were obtained for 12 asteroids, namely (122) Gerda, (152) Atala, (260) Huberta, (665) Sabine, (692) Hippodamia, (723) Hammonia, (745) Mauritia, (768) Struveana, (863) Benkoela, (1113) Katja, (1175) Margo, and (2057) Rosemary. For five of them, we have obtained the magnitude-phase relations in the region of the opposition effect (OE) down to phase angles < 1 deg. For an estimation of the absolute magnitudes of the asteroids, we used the new $HG_1G_2$ – magnitude system proposed by Muinonen et al. (2010), with some modifications presented by Penttilä et al. (2016). For actual computations, the online calculator for the $HG_1G_2$ photometric system (http://h152.it.helsinki.fi/HG1G2/) was used. Amplitudes of OE were determined using the recommendation of Belskaya and Shevchenko (2000).

The aspect data of the observed asteroids and the measured magnitudes at the lightcurve primary maximum, together with their accuracy, are given in Table 1. In Table 2, we present the results of our observations: rotation periods, lightcurve amplitudes, the average color indices *B-V* and *V-R,* and the *H*, $G_1$, $G_2$ parameters of the magnitude-phase function. As it was pointed out by Zappala et al. (1990) and was reproduced in the numerical computations by Muinonen (1998), the asteroid lightcurve amplitude increases with increasing phase angle. We determined absolute magnitude $H$, and $G_1$, $G_2$ parameters and linear phase coefficient $\beta$ for the lightcurve primary maximum to correct the phase angle effects. In the same table, we also give semi-major axes of the orbit of the observed asteroids, their taxonomic class, albedo, and diameter with the corresponding references.

Below, we give brief characteristics for the observed asteroids.

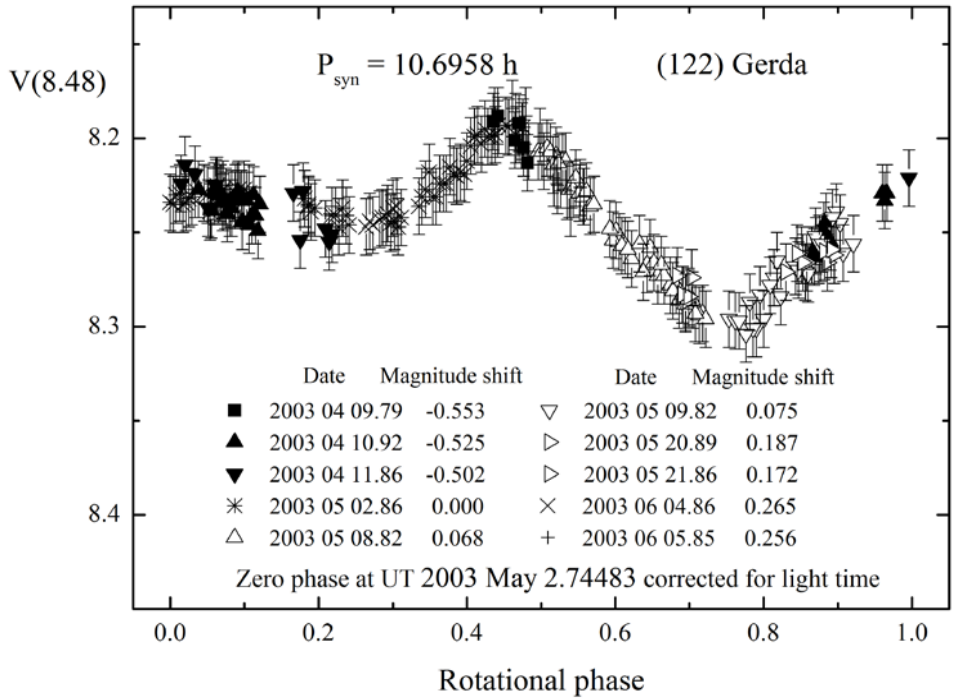

Fig. 1. Composite lightcurve of asteroid (122) Gerda.

*(122) Gerda.* The asteroid has an orbit with a semi-major axis of 3.23 au, an eccentricity of 0.03, an inclination of 1.6°, and it does not belong to any asteroid family. It is a relatively large asteroid with a diameter of about 85 km (Usui et al., 2011), and surface albedo ranging from 0.17 (Usui et al., 2011) up to 0.30 (Masiero et al. 2012). From the visible spectrum, Bus and Binzel (2002) classified this object as an L-type. Devogèle et al. (2018), from the visible and near infrared spectrum, attributed it to S-asteroids, and Belskaya et al. (2017) from the polarimetric data assumed that it is a K-type.

The photometric observations for shape reconstruction were performed by many authors (e.g. Devogèle et al., 2017; Hanuš et al., 2016; Shevchenko et al., 2009). Martikainen et al. (2021) using the model developed by Muinonen et al. (2020) have performed modelling and have also obtained shape model, pole coordinates, and phase curve parameters from Gaia photometry for this asteroid. However, the quality magnitude-phase relation including small phase angles was not obtained for this asteroid. Our observations were performed on ten nights in April-June 2003 (see Table 1) covering phase angles down to 0.3°. The composite lightcurve constructed with the period $10^h.6958 \pm 0.0005$ is shown in Fig.1. The measured period is slightly different from the periods obtained by Hanuš et al. (2016) ($10^h.68724$), Devogèle et al. (2017) ($10^h.6872$) and Martikainen et al. (2021) ($10^h.6872$). The magnitude-phase relation corrected for lightcurve variations is presented in Fig. 2. The solid line indicates the approximation of the phase curve by the $HG_1G_2$-function (Muinonen et al., 2010; Penttilä et al., 2016) with the parameters listed in Table 2.

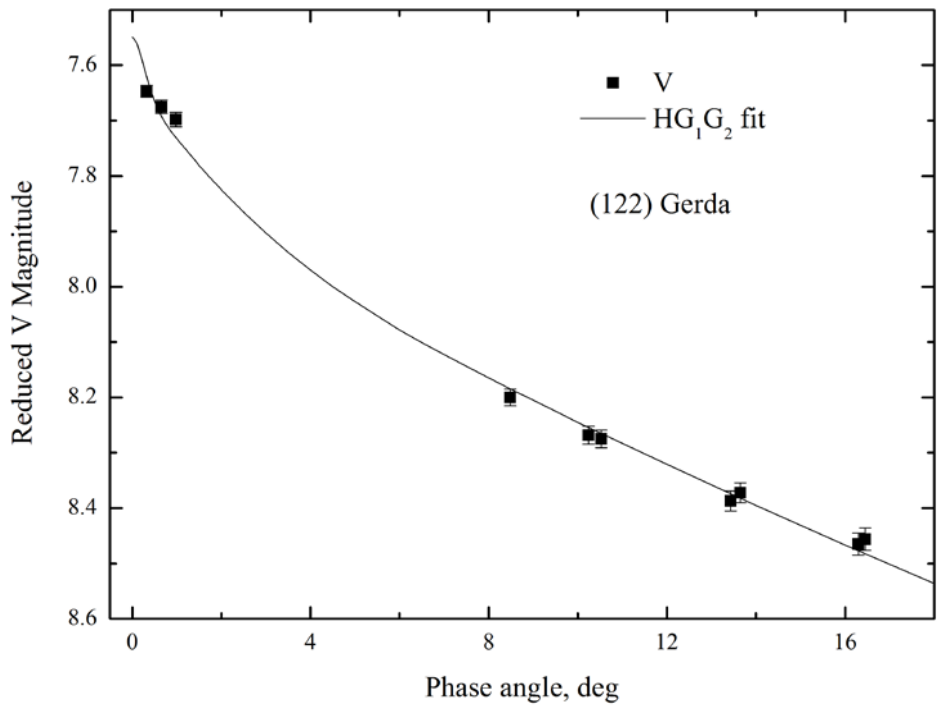

Fig. 2. Magnitude-phase relation of asteroid (122) Gerda (solid line is the best fit with the $HG_1G_2$-function).

The linear phase coefficient $\beta$ is equal to $0.033 \pm 0.001$ mag/deg, and the amplitude of the OE is $0.30 \pm 0.02$ mag. Taking into account the lightcurve amplitude, the rotationaly averaged absolute magnitude is equal to $7.60 \pm 0.02$ mag. For comparison with the Minor Planet Center (hereafter MPC), we present also the parameters of the *HG*- function (Bowell et al., 1989): $H = 7.65 \pm 0.02$ mag, $G = 0.15 \pm 0.02$. There are small differences in the absolute magnitudes obtained with our data using different phase functions, but the value adopted by the MPC with the *HG*- function ($H=7.71$) is much bigger. On the other hand, our value is close to the one obtained by Veres et al. (2015) ($H=7.58$). We have also measured color indices for this asteroid $B - V = 0.88 \pm 0.02$ and $V - R = 0.45 \pm 0.02$ mag, that are typical for S-asteroids (Lupishko et al. 2007, Shevchenko and Lupishko 1998).

*(152) Atala.* This asteroid has an orbit with an eccentricity of 0.079 and a semi-major axis of 3.14 au. The albedo values provided for this asteroid by the WISE and AKARI IR surveys are 0.24 and 0.27, with the diameters of 61 km and 57 km, respectively (Masiero et al. 2014, Usui et al. 2011). The visible reflectance spectra of this asteroid has a slope of 10.6 %/$10^3$ Å and an absorption band at about 0.90 μm, which characterizes a substance consisting predominantly with pyroxenes. Because of that it was classified as an S-asteroid (Bus and Binzel 2002). However, the polarimetric properties of (152) Atala differ from typical ones for the S-type (Fornasier et al., 2006), and are closer to the A-type (Belskaya et al. 2017). The diameter of Atala derived from the stellar occultation is equal to 66.4 km (Dunham et al. 2016). To estimate the albedo using this diameter, the correct determination of the absolute magnitude at the time of the occultation is needed. Previous photometric observations of (152) Atala, with a goal of determination of its rotational properties, were performed by Schober (1983), Durech et al. (2011), Hanuš et al. (2013b), and Behrend et al. (2020).

We carried out new photometric observations of this object on eleven nights in April-June 2017 to measure the magnitude-phase dependence. We were able to cover both the linear region of the magnitude-phase curve, and the area of the opposition effect down to 1.4 degree of the phase angle. The lightcurve amplitude (Fig. 3) is equal to 0.30 mag and the rotation period is $6^h.2447 \pm 0.0002$. The later coincides with periods obtained by Durech et al. (2011), and Hanuš et al. (2013b).

The phase curves of (152) Atala in the *V* and *R* bands are shown in Fig. 4. They were corrected for the lightcurve amplitude effect. We have not found any noticeable differences between the phase curves in the *V* and *R* bands except for the vertical shift equal to $V - R = 0.49$ mag in all measured phase angles. The solid line represents the fit of the *V*-band phase curve with the $HG_1G_2$-function with the parameters: $H = 8.055 \pm 0.015$ mag, $G_1 = 0.30 \pm 0.03$, $G_2 = 0.29 \pm 0.02$. The absolute magnitude corrected for the lightcurve amplitude (8.20) is different from the value given by the MPC ($H = 8.31$). The linear phase coefficient $\beta$ ($0.032 \pm 0.002$ mag/deg) and the amplitude of the opposition effect ($0.37 \pm 0.04$ mag) have typical values for the moderate albedo asteroids.

Our observations in 2017 were carried out at a similar aspect as in 2006, when the occultation diameter of (152) Atala was measured (Dunham et al. 2016). Lightcurve obtained shortly after the occultation and published by Behrend et al. (2020) was used by us to determine the absolute magnitude at that moment.

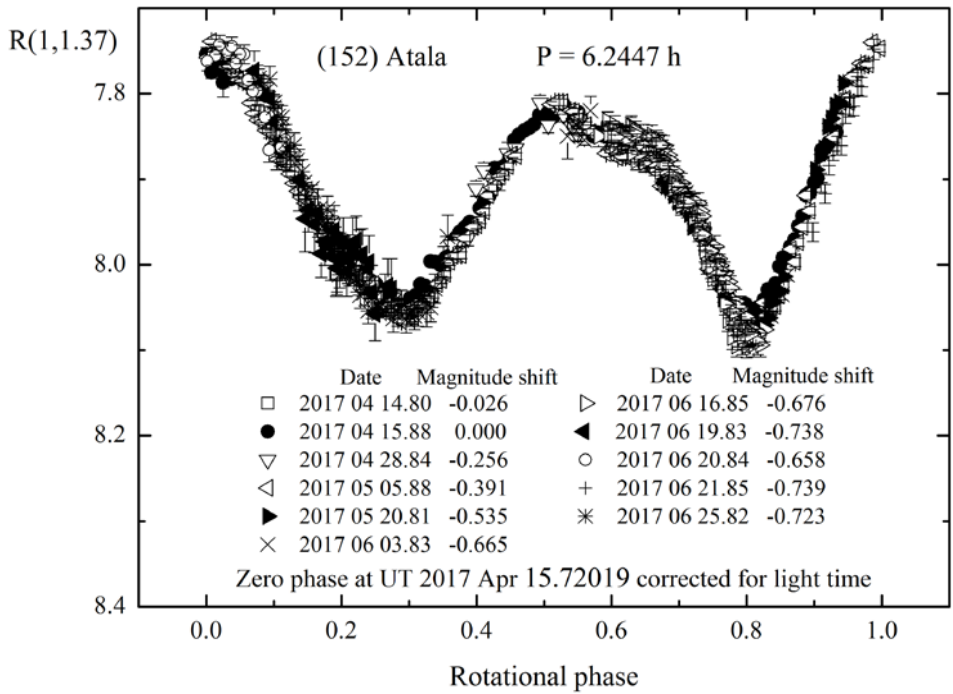

Fig. 3. Composite lightcurve of asteroid (152) Atala.

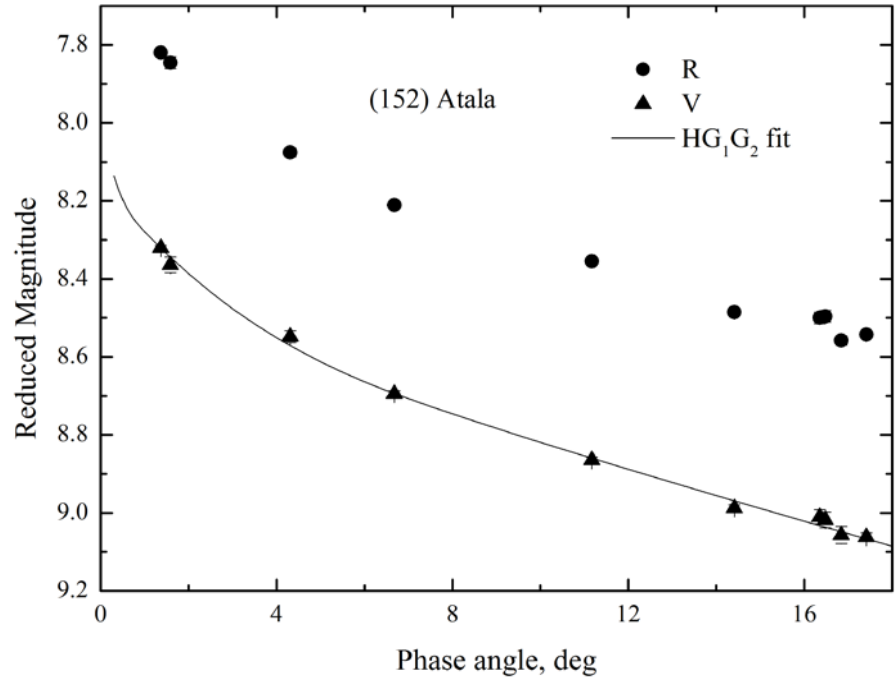

Fig. 4. Magnitude-phase relation of asteroid (152) Atala (solid line is the best fit with the $HG_1G_2$-function).

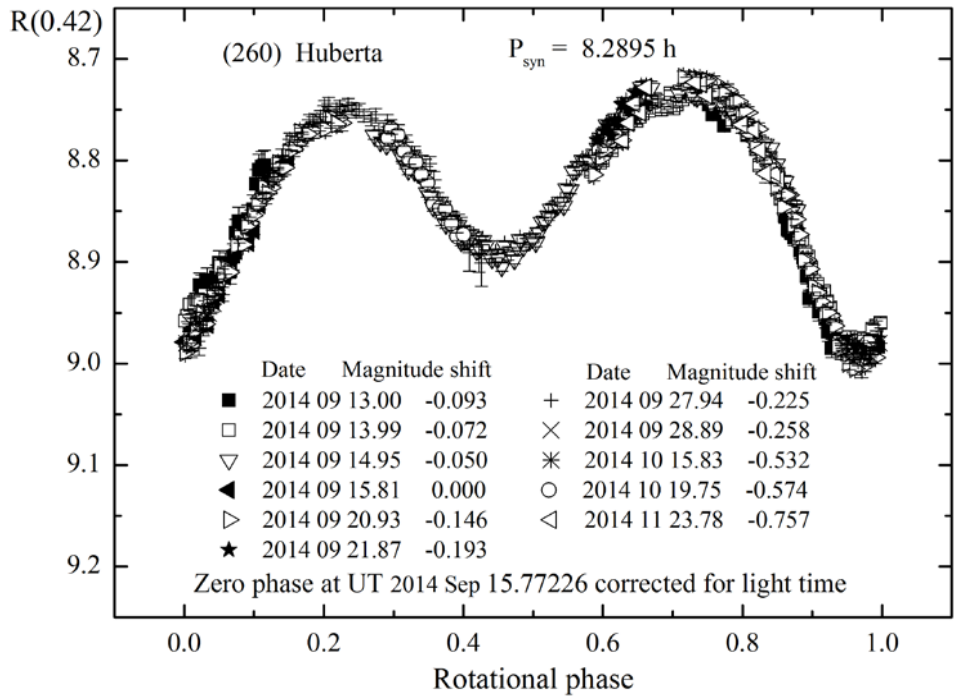

Fig. 5. Composite lightcurve of asteroid (260) Huberta.

We found that the difference *Δm* between relative *V* magnitudes at the maximum of the lightcurve and at the moment of the occultation was 0.20 ± 0.02 mag. Adding *Δm* to the value of the absolute magnitude *H* obtained in this work, we calculated albedo of (152) Atala and got the value of 0.20 ± 0.02.

*(260) Huberta.* This asteroid orbits around the Sun with a semi-major axis of 3.45 au and belongs to the Cybele group. It is a low-albedo object with a diameter of about 100 km (Masiero et al., 2014; Usui et al., 2011). Albedo values of this asteroid from different determinations are in the range from 0.044 to 0.054 (Masiero et al., 2014; Usui et al., 2011). We used this body as a control object to show real differences in magnitude-phase relations between moderate and low albedo asteroids (see Section 3). Lagerkvist et al. (2001) performed the first photometric observations of this object during four nights to obtain its rotation period ($8^h.29$). More intensive observations for shape reconstruction were performed by Hanuš et al. (2013a), but the magnitude-phase relation was not obtained so far.

Our observations were performed on eleven nights in September-November 2014 (see Table 1) in the range of phase angles from 18° down to 0.3°. The composite lightcurve constructed with the rotation period $8^h.2895 \pm 0.0005$ is presented in Fig. 5 and the magnitude-phase relation is shown in Fig. 6. The solid line in Fig. 6 is approximation with the $HG_1G_2$-function. The obtained values of $G_1$ and $G_2$ parameters (see Table 2) are close to average values for the low albedo asteroids (Shevchenko et al. 2016). It should be noted that our estimation of the mean absolute magnitude (*H*=9.12) has a slight difference with that given by the MPC (*H*=9.18) and differs from obtained by Veres et al. (2015) (*H*=8.88). The linear phase coefficient *β* is equal to 0.040 ± 0.001 mag/deg, and the amplitude of the OE is 0.17 ± 0.02 mag in the *V* band.

*(665) Sabine.* The asteroid has an orbit with a semi-major axis of 3.14 au, an eccentricity of 0.17, and an inclination of 14.8°. Although it does not belong to any family, in the proper elements space, it is very close to the Tirela family. According to the IRAS data, its albedo is equal to 0.39 with a diameter of about 51 km (Tedesco et al. 2002). The high albedo was also confirmed by the AKARI data (0.36, Usui et al. 2011). The data from the WISE satellite give even higher albedo value of 0.49 (Masiero et al. 2012). In addition, the reflectance spectrum of the asteroid's surface has a slope of only 3% / $10^3$ Å in the visible range without noticeable absorption bands. Usually it was linked to E-type asteroids (Tholen 1989; Bus and Binzel 2002; DeMeo et al., 2009), although Lazarro et al. (2004) classified it as the X-type object.

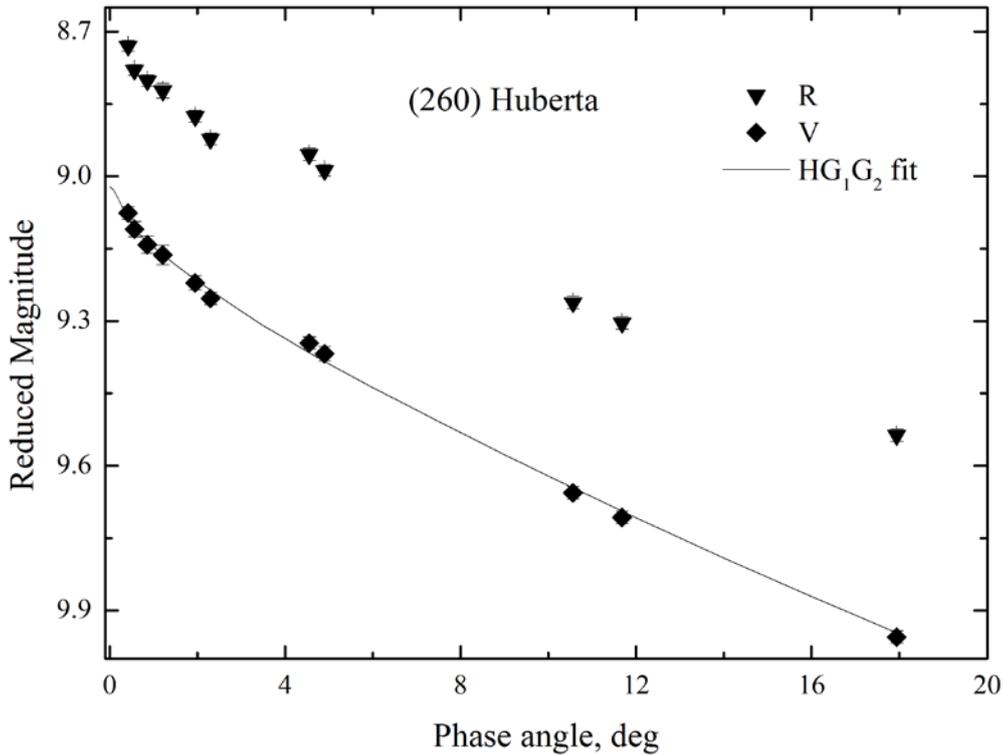

Fig. 6. Magnitude-phase relation of asteroid (260) Huberta (solid line is the best fit with the $HG_1G_2$-function).

Previous photometric observations of this asteroid were carried out to determine its rotational characteristics and shape (Michalowski et al. 2006), but the magnitude-phase dependence was not previously obtained.

Our observations were performed on twenty-three nights in January-May 2018 (see Table 1). We used the *BVR* bands and covered both the linear region and the area of the opposition effect down to the phase angle of 1°. The preliminary results were presented in Shevchenko et al. (2019b). The composite lightcurve with a rotation period of $4^h.29407$ is shown in Fig. 7. The obtained period is close to the values obtained by Behrend et al. (2020), Michalowski et al. (2006) and Pal et al. (2020). The maximum amplitude of the lightcurve is found to be 0.35 mag. The lightcurve variations were taken into account in the phase dependence of brightness. We have also obtained the average values of the color indices $B - V = 0.69$ and $V - R = 0.40$ mag.

The phase curves for the maximum brightness of (665) Sabine in the $V$ and $R$ bands are shown in Fig. 8. The solid line indicates the approximation of the phase curve by $HG_1G_2$-function with the parameters: $H = 8.50$ mag, $G_1 = 0.52$, $G_2 = 0.19$ (see Table 2). Taking into account the lightcurve amplitude, our estimate of the absolute magnitude is 8.66 mag and that is close to the value given by the MPC ($H=8.71$). The linear phase coefficient $\beta$ for the $V$ band is equal to $0.037 \pm 0.001$ mag/deg, and the amplitude of the OE is about 0.26 mag. These values are atypical for a high albedo surface and exclude the E-type classification for this asteroid.

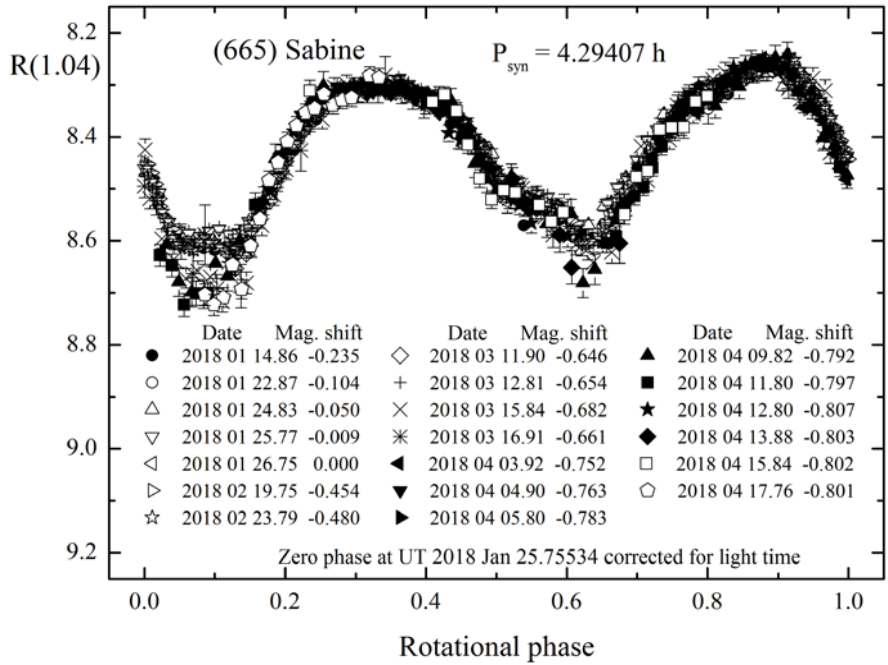

Fig. 7. Composite lightcurve of asteroid (665) Sabine.

*(692) Hippodamia*. This outer main belt asteroid has a high inclination of the orbit equal to 26°.1, a semi-major axis of 3.38 au, the eccentricity of 0.17 and belongs to the Cybele group. It was classified as a taxonomical class S (Tholen 1989). According to the infrared data of the IRAS, WISE and AKARI satellites, its diameter is about 45 km, but albedo varies in the range from 0.18 to 0.24 (Ali-Lagoa et al., 2018; Masiero et al., 2014; Tedesco et al., 2002; Usui et al., 2011). For the first time photometric observations of the Hippodamia were performed by Zappala et al. (1989), who determined the rotation period of $8^h.98$ with the amplitude of the lightcurve of 0.50 mag. Other observers have also carried out photometrical observations to measure spin rate of this asteroid (Behrend et al., 2020; Hanus et al., 2011, 2016; Pal et al. 2020).

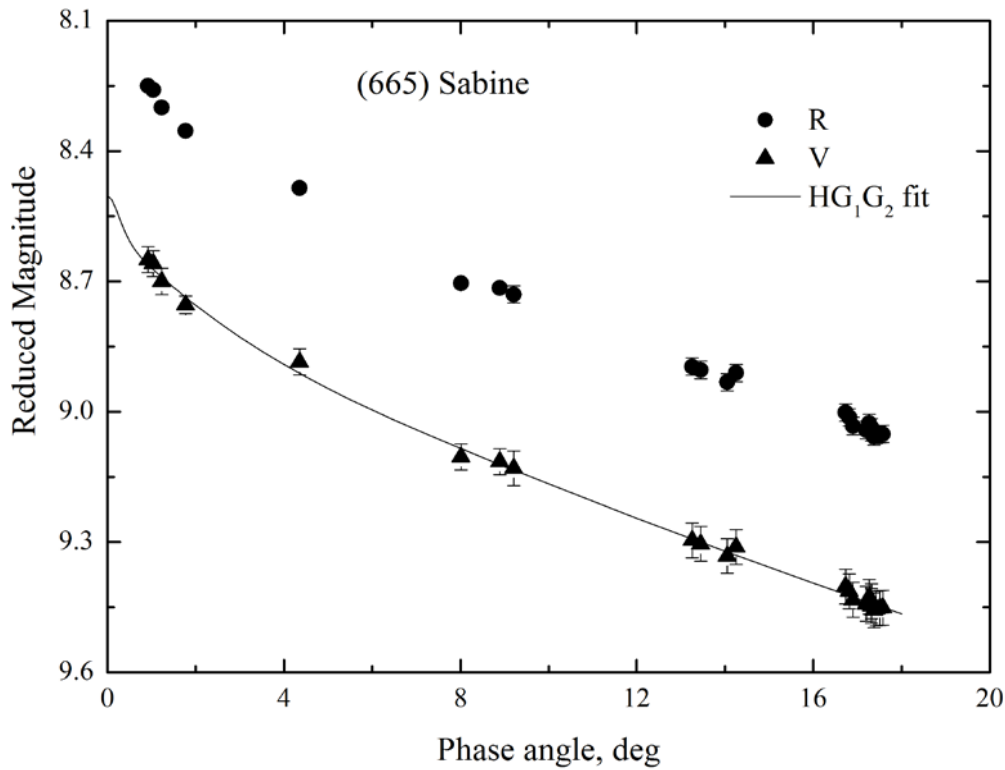

Fig. 8. Magnitude-phase relation of asteroid (665) Sabine (solid line is the best fit with the $HG_1G_2$-function).

Our observations were carried out for only four nights (see Table 1) to obtain the magnitude-phase relation. As a result, the composite lightcurve was obtained with the rotation period of $8^h.9977$ (Fig. 9). This value is close to the values obtained by Behrend et al. (2020), Hanus et al. (2016) and Pal et al. (2020). Although, our observations do not cover the full lightcurve (see Fig. 9), the light curve amplitude is well-determined (0.32 mag). The average color index is $V-R$ = 0.48 mag.

Our observations covered the linear part of the magnitude phase curve only. The obtained $V$ and $R$ band phase curves are presented in Fig. 10. The linear phase coefficient $\beta$ for the $V$ band is equal to 0.022 ± 0.001 mag/deg, which is inherent for a high albedo surface. Since we had only four values of magnitudes on the phase curve, we did not fit them of the three parameter $HG_1G_2$-function. To obtain the absolute magnitude, we used the average values of $G_1$ and $G_2$ derived for high-albedo asteroids by Shevchenko et al. (2016). The absolute magnitude obtained in this way is $H$ = 9.23 ± 0.12 mag and the amplitude of OE is estimated to be 0.21.

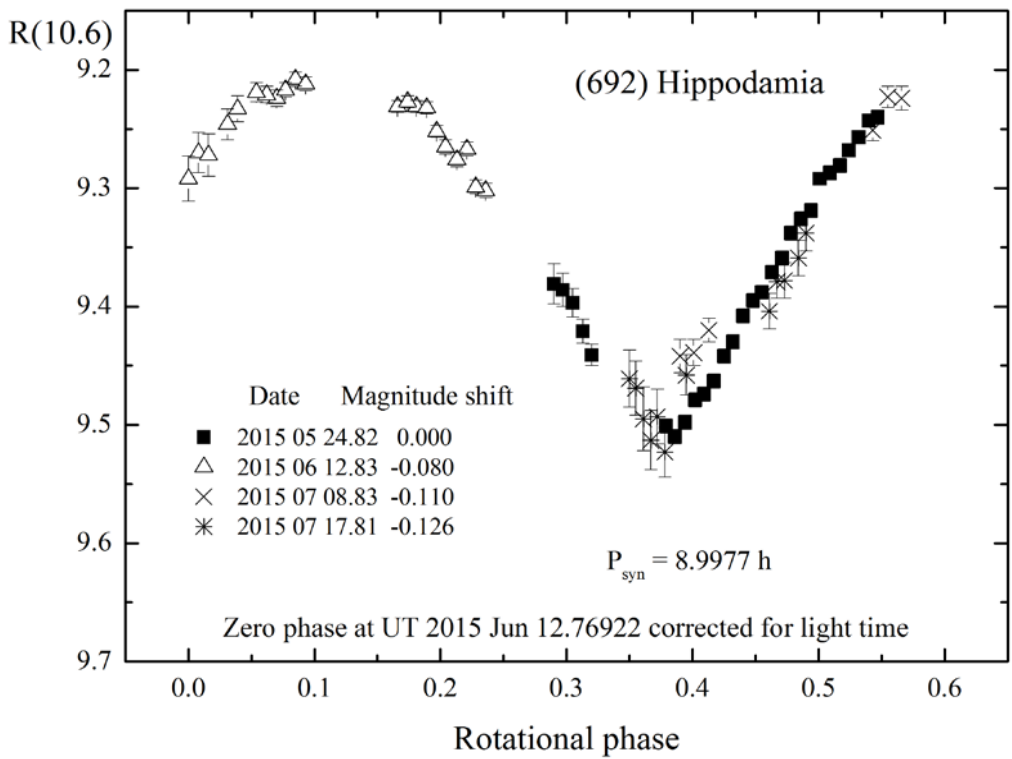

Fig. 9. Composite lightcurve of asteroid (692) Hippodamia.

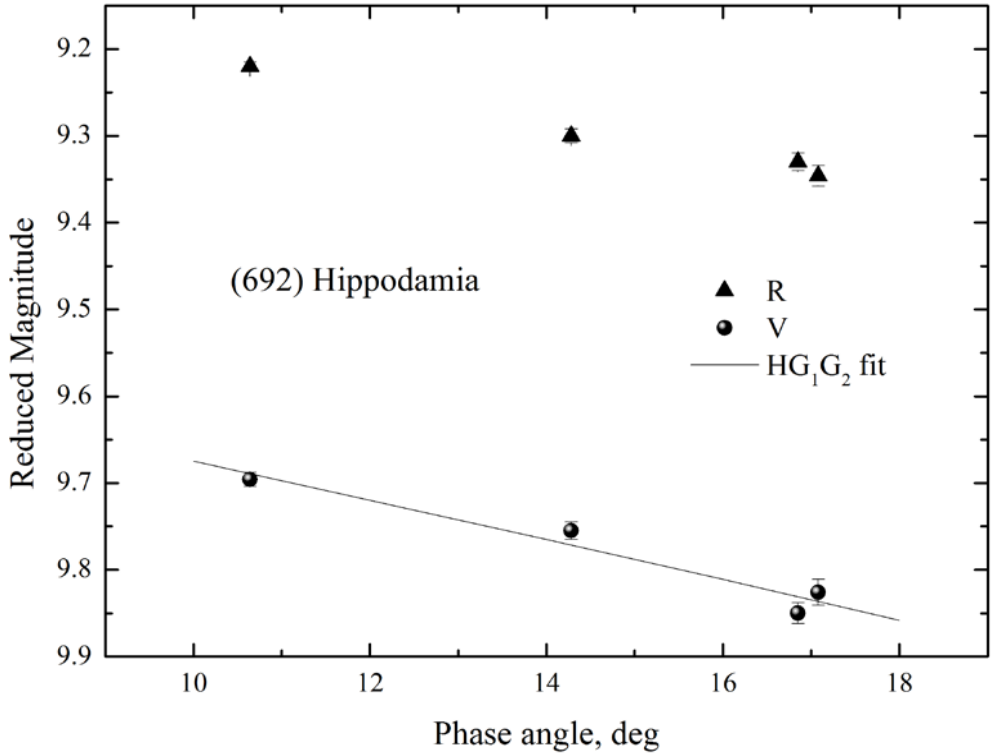

Fig. 10. Magnitude-phase relation of asteroid (692) Hippodamia (solid line is the best fit with the $HG_1G_2$-function).

For comparison with the MPC data, we calculated also parameters of the *HG*- function: $H = 9.26 \pm 0.09$ mag, $G = 0.48 \pm 0.11$. The former is greater than the value given by the MPC ($H=9.02$) even without taking into account the amplitude of the lightcurve. This, however, can be attributed to the lower accuracy of our *H* value.

*(723) Hammonia*. The asteroid has an orbit with a semi-major axis of 2.99 au and a small eccentricity (0.06). Initially (723) Hammonia was classified by Bus and Binzel (2002) as a C-type asteroid, but estimations of its albedo obtained from WISE (Masiero et al. 2012), AKARI (Usui et al. 2011), and IRAS (Tedesco et al. 2002) are equal to 0.35, 0.29, and 0.18, respectively, and contradict this classification. Lightcurve data for this asteroid were obtained by Behrend et al. (2020) who determined the rotation period ($5^h.436$), and Durech and Hanus (2018) who determined pole coordinates. We observed this asteroid for seventeen nights in 2014 to obtain a good quality magnitude-phase relation. The preliminary results were presented in Shevchenko et al. (2015). We obtained the composite lightcurve with the rotation period of $5^h.4348 \pm 0.0015$ and the maximum amplitude of $0.08 \pm 0.02$ mag (Fig. 11). The average color index $V$-$R$ is equal to 0.35 mag. The magnitude-phase relation of (723) Hammonia in $V$ and $R$ bands corrected for lightcurve amplitude is shown in Fig. 12. The solid line indicates the approximation of the phase dependence by the $HG_1G_2$-function with the parameters: $H = 9.88$ mag, $G_1 = 0.24$, $G_2 = 0.43$ (see Table 2). It should be noted that our estimates of absolute magnitude are close to the value given by Veres et al. (2015) ($H$=9.90) and differ from the MPC ($H$=10.04). The linear phase coefficient $\beta$ for the $V$ band is equal to $0.027 \pm 0.002$ mag/deg, and the amplitude of the OE is about 0.31 mag.

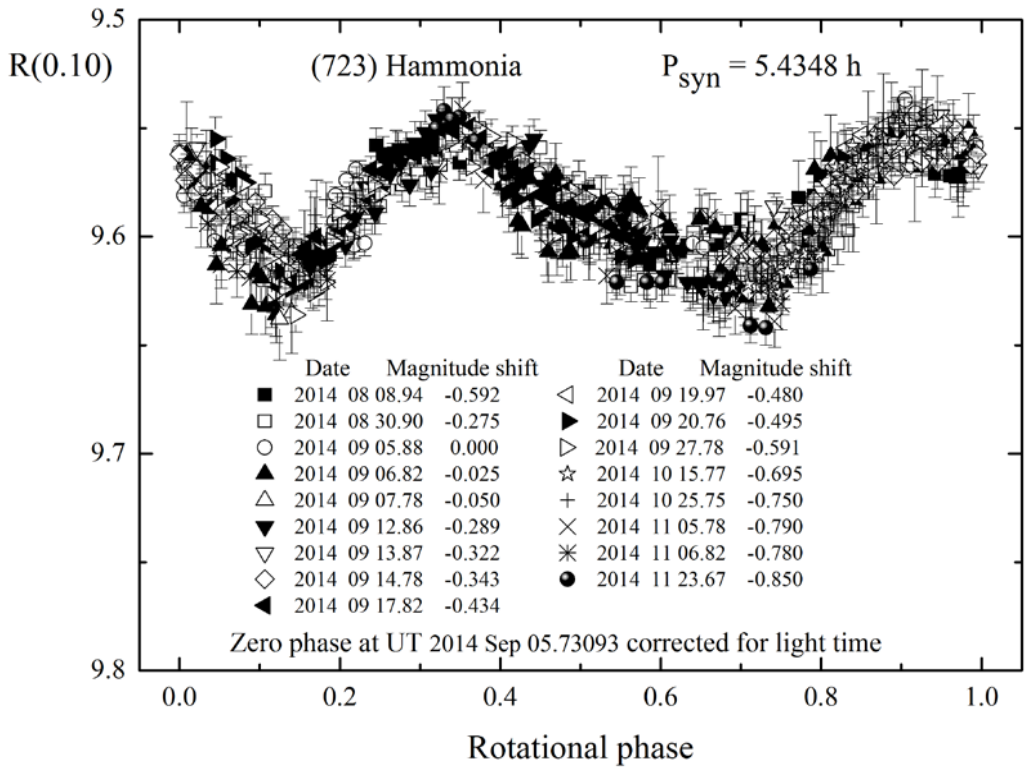

Fig. 11. Composite lightcurve of asteroid (723) Hammonia.

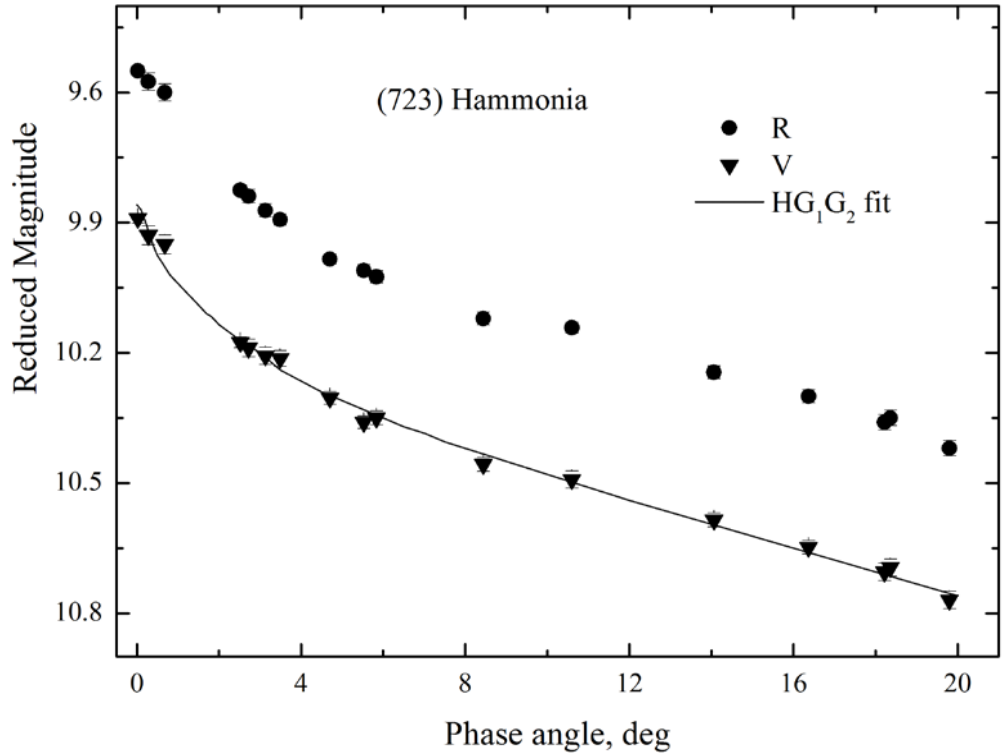

Fig. 12. Magnitude-phase relation of asteroid (723) Hammonia (solid line is the best fit with the $HG_1G_2$-function).

*(745) Mauritia.* The asteroid orbits in the outer part of the main belt with a semi-major axis of 3.26 au, an eccentricity of 0.04 and an inclination of 13°.3. It has a diameter about 25 km with the albedo ranging from 0.20 to 0.25 (Mainzer et al., 2016; Masiero et al., 2014; Usui et al., 2011). Pilcher (2013) observed this asteroid in 2013 and obtained a rotation period of $9^h.945$. Our observations were carried out for five nights in 2018, in the *V* and *R* bands, with a phase angle changing from 2°.8 to 18°.6.

A composite lightcurve with rotation period of $9^h.9425$ is presented in Fig. 13. Our value of the rotation period is more precise than that obtained by Pilcher (2013). The maximum amplitude of the lightcurve for this opposition is 0.37 ± 0.02 mag, and the average color index is *V-R* = 0.42 mag. Our data made it possible to obtain the magnitude– phase relation in two bands (Fig. 14). As in the case of asteroid (692) Hippodamia, we have not enough data to obtain accurate values of parameters of the $HG_1G_2$-function.

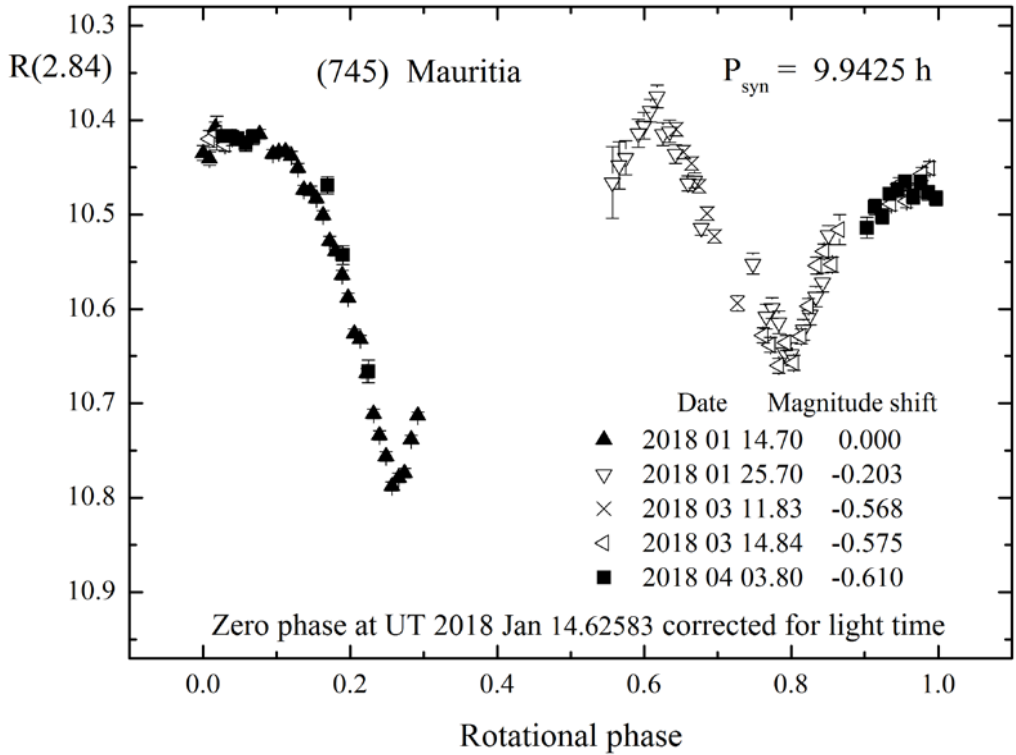

Fig. 13. Composite lightcurve of asteroid (745) Mauritia.

The values of all three parameters (see Table 2) have rather large errors. It should be noted, that this function inside the measured phase angle range gives correct values of magnitudes. There is a problem with the extrapolation down to the zero phase angle; the function overestimates the absolute magnitude $H$ (10.43). For comparison, we present also parameters of the $HG$- function for $V$ band: $H = 10.53 \pm 0.02$ mag, $G = 0.21 \pm 0.03$. In this case, the value of the absolute magnitude is more reliable. The parameters of the phase functions and the estimate of the linear phase coefficient (0.032 mag/deg) indicate a moderate albedo of this asteroid. Our estimation of the rotationally averaged absolute magnitude is 10.72 mag, which is different from the value given by the MPC ($H$=10.54) and by Veres et al. (2015) ($H$=10.33).

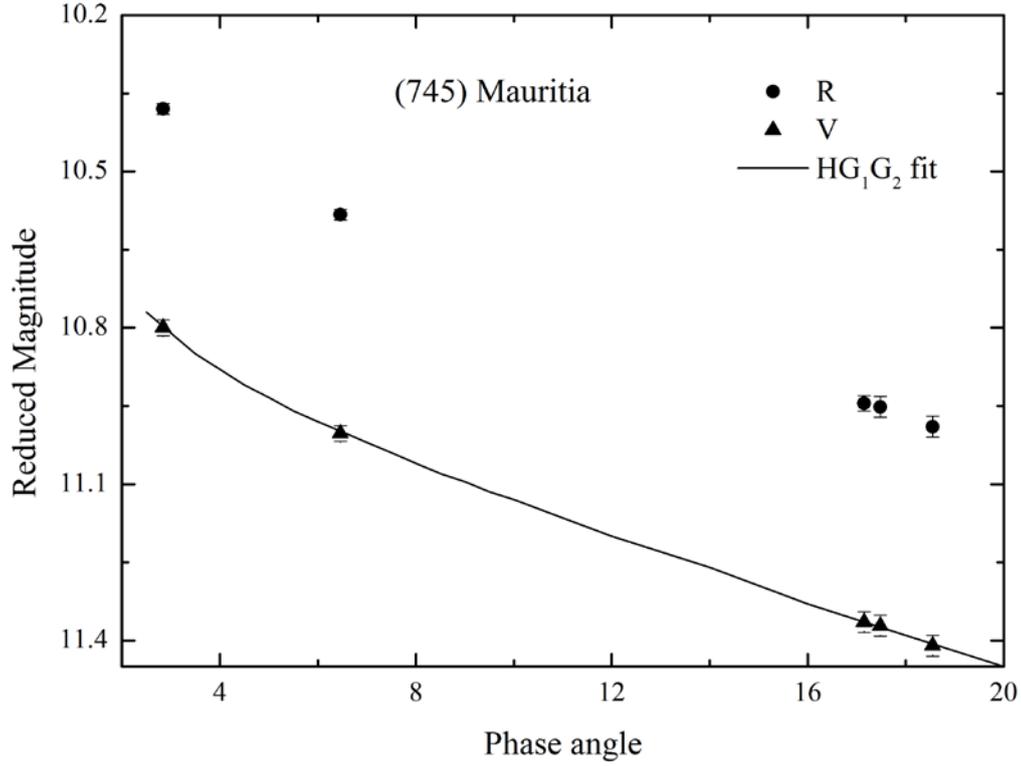

Fig. 14. Magnitude-phase relation of asteroid (745) Mauritia (solid line is the best fit with the $HG_1G_2$-function).

*(768) Struveana*. The asteroid has a semi-major axis of 3.14 au and the eccentricity of 0.21. It belongs to the asteroid family of (137) Meliboea (Nesvorny 2015). Based on its spectrum, this asteroid was classified as an X- type (Tholen, 1989; Lazarro et al., 2004). According to the infrared data obtained by the WISE and AKARI satellites, it has a dimeter of about 30 km and the albedo from 0.14 to 0.24 (Ali-Lagoa et al., 2018; Masiero et al., 2011; Usui et al., 2011). Its rotation period ($10^h.76$) was obtained by Gil-Hutton et al. (2003).

The asteroid was observed for nine nights during three months in 2016 (Table 1), which made it possible to determine the rotation period more precisely: $10^h.7458 \pm 0.0005$. A composite lightcurve with the lightcurve amplitude of 0.20 mag is shown in Fig. 15. The phase curve of this asteroid is presented in Fig. 16 in the *R* filter. A solid line indicates a fit of the phase curve with the $HG_1G_2$- function with the parameters: $H_R = 9.409$ mag, $G_1 = 0.19$, $G_2 = 0.29$ (see Table 2). To transform the $H$ value from the $R$ band to the $V$ band we used the average value of $V-R = 0.45$, which has been derived for the moderate albedo asteroids by Shevchenko and Lupishko (1998). As a result we obtained $H = 9.96$ mag, which is different from the value given by the MPC ($H = 10.28$) and slightly different from that of Veres et al. (2015) ($H = 9.90$). The linear phase coefficient $\beta$ for the $R$ band is equal to $0.033 \pm 0.001$ mag/deg, and the amplitude of the OE is about 0.33 mag (using the $HG_1G_2$-function for approximation to the zero phase angle).

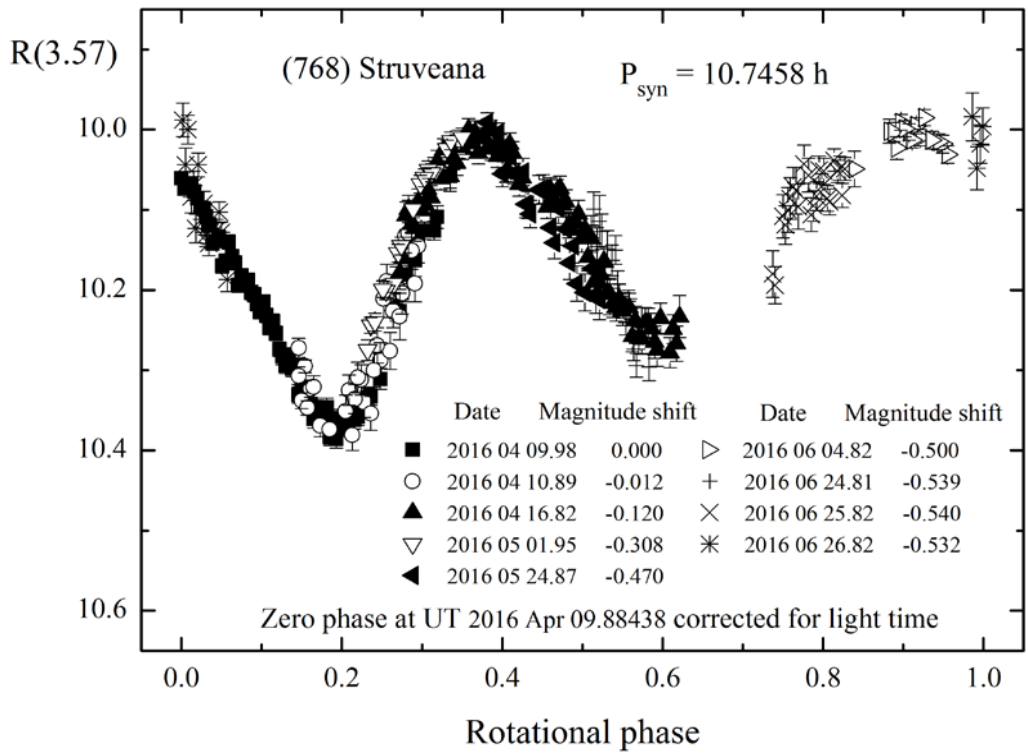

Fig. 15. Composite lightcurve of asteroid (768) Struveana.

*(863) Benkoela.* The asteroid has a nearly circular orbit with an eccentricity of 0.03, inclination of 25.4° and a semi-major axis of 3.20 au. It does not belong to any asteroid family. In the proper element space, it lies in the low asteroid density region, at smaller inclinations than the Euphrosyne family. Morrison and Chapman (1976) made the first albedo estimation (0.26) from ground-based radiometric measurements and found its similarity to moderate albedo asteroids. Other data on the albedo obtained from IRAS (0.60, Tedesco et al., 2002), WISE (0.79, 0.29, Masiero et al., 2012, 2014) and AKARI (0.44, Usui et al., 2011) pointed out possible high albedo surface of this body. The asteroid was classified as an A-type (Bus and Binzel 2002; DeMeo et al., 2009; Tholen 1989; Xu et al. 1995). It should be noted that the asteroids of A-type have typically moderate albedo. The polarimetric properties of (863) Benkoela are also close to the A-type (Belskaya et al. 2017, Lopez-Sisterna, et al. 2019) and the estimated polarimetric albedo is 0.32 ± 0.12 (Fornasier et al. 2006). Benkoela is the only object of this type in the outer part of the main belt.

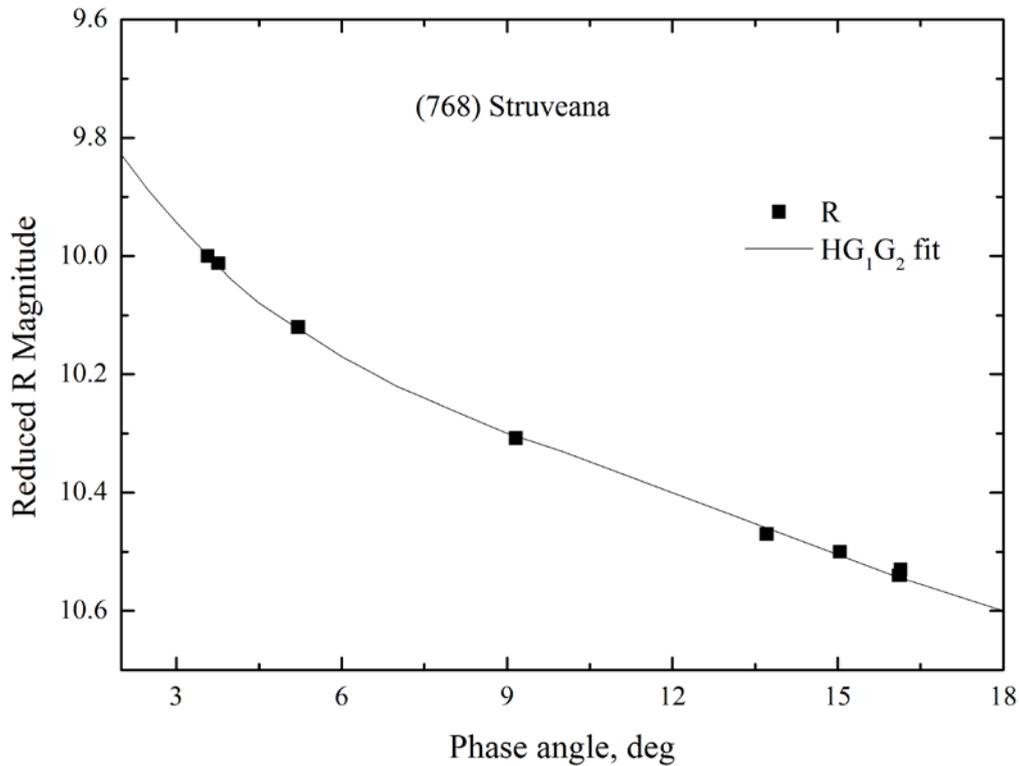

Fig. 16. Magnitude-phase relation of asteroid (768) Struveana (solid line is the best fit with the $HG_1G_2$-function).

Previous photometric observations of (863) Benkoela (Garceran et al., 2016; Harris et al., 1989; Warner 2004) were performed for determination of its rotational properties. We made new photometric observations of the asteroid on eight nights in October-November 2018 and January 2019. The preliminary results were presented in Shevchenko et al. (2019b). We were able to cover only the linear part of the magnitude-phase curve down to 9.4 degree of the phase angle. We were not able to determine the rotation period because of the small (0.05 mag, see Fig. 17) amplitude of the lightcurve. For successful determination of the period, new observations at other apparitions are required. From the collected data we managed to derive the average values of the color indices, *B-V* = 1.06 and *V-R* = 0.56 mag, that suggested a very high spectral slope. Our obtained value of *B-V* color index is close to that in Tedesco (1989) (*B-V* = 1.08 mag).

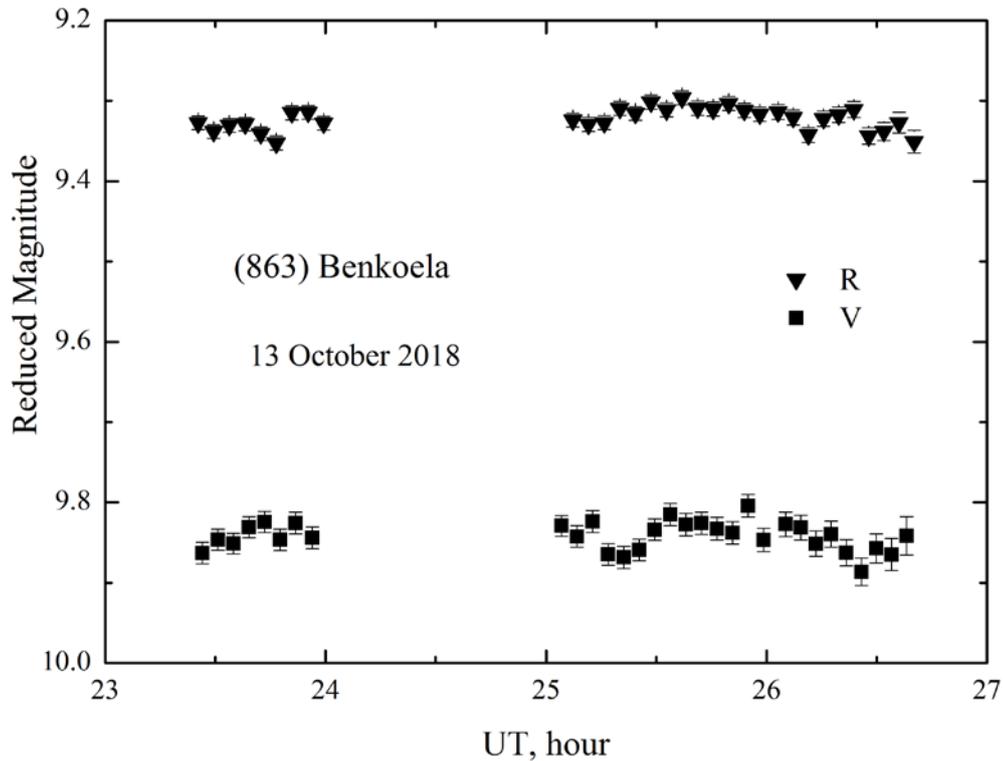

Fig. 17. Lightcurve of asteroid (863) Benkoela on 13 October 2018.

Since the lightcurve amplitude of Benkoela is small, we used an average magnitude from a night to construct the magnitude-phase curve (Fig. 18). The solid line represents the approximation of the phase curve of the $HG_1G_2$-function with the assumed average parameters of $G_1 = 0.26$ and $G_2$

= 0.37 for moderate albedo asteroids (Shevchenko et al. 2016). The fitted value of the absolute magnitude was $H = 9.16 \pm 0.58$ mag. The linear slope of the magnitude-phase curve of (863) Benkoela is found to be $\beta = 0.034 \pm 0.004$ mag/deg. Up to now there are no well measured magnitude-phase relations for A-type asteroids. Because of that, we plan to continue the observations of this asteroid to cover the opposition effect region.

*(1113) Katja*. This asteroid belongs to the outer main belt and has a semi-major axis of 3.11 au, an eccentricity of 0.14 and an inclination of 13.3°. The size of this asteroid is about 40 km with an albedo from 0.11 to 0.25 (Ali-Lagoa et al., 2018; Masiero et al., 2014; Usui et al., 2011; Tedesco et al., 2002). The photometric observations to obtain the rotational properties were carried out by Behrend et al. (2020), Pal et al. (2020) and Robinson et al. (2002). We observed Katja on ten nights in 2016 and obtained a composite lightcurve with period of $18^h.432$ and with a maximum amplitude of 0.12 mag (Fig. 19). Our value of the rotation period is close to that obtained by Pal et al. (2020).

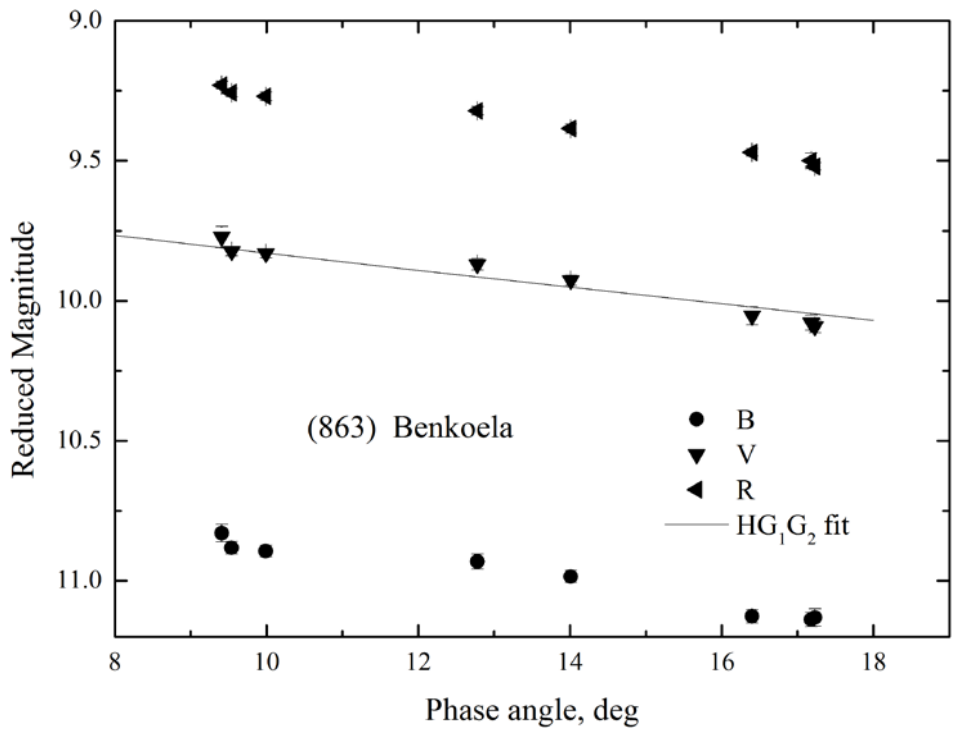

Fig. 18. Magnitude-phase relation of asteroid (863) Benkoela (solid line is the best fit with the $HG_1G_2$-function).

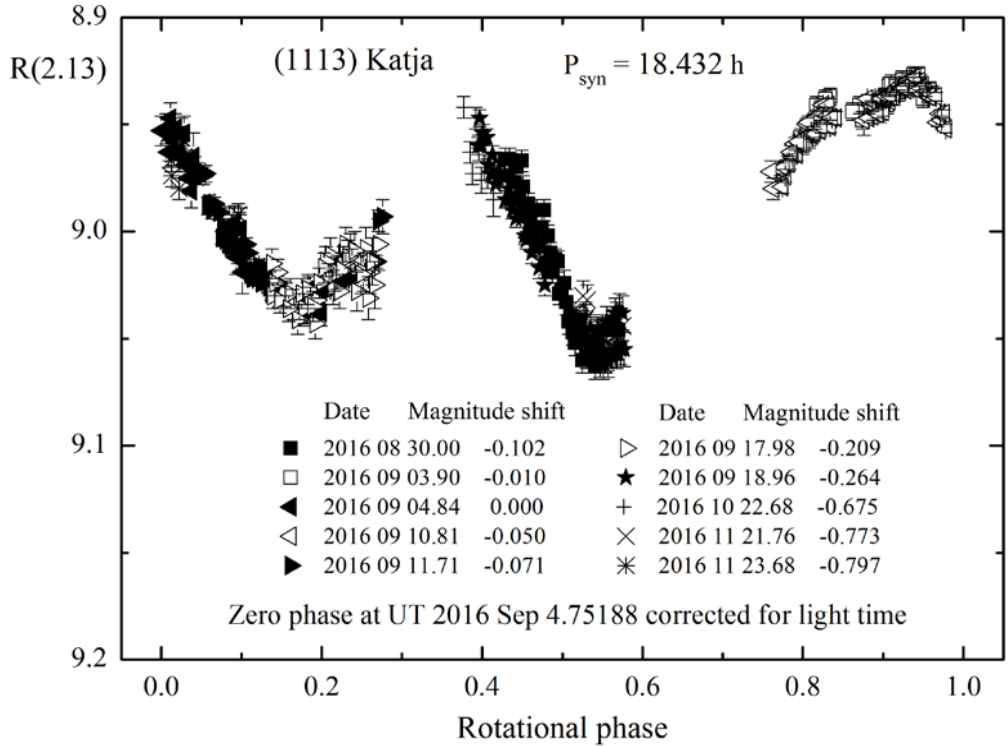

Fig. 19. Composite lightcurve of asteroid (1113) Katja.

The phase curves of Katja in *V* and *R* bands are presented in Fig. 20. The solid line indicates the approximation of the phase curve with the $HG_1G_2$-function with the parameters $H = 9.05$ mag, $G_1 = 0.36$, $G_2 = 0.20$ (see Table 2). Our estimate of the absolute magnitude differ from the value given by Veres et al. (2015) ($H=9.49$) and by the MPC value ($H=9.35$). The linear phase coefficient $\beta$ that was estimated for the *V* band is equal to $0.035 \pm 0.002$ mag/deg, and the amplitude of the OE is about 0.31 mag.

*(1175) Margo.* The asteroid has a semi-major axis of 3.21 au, an eccentricity of 0.07, an inclination of 16°.3, and belongs to the asteroids of the outer part of the main belt. This asteroid was classified as a taxonomic class S (Bus, Binzel 2002). According to data obtained by WISE and AKARI satellites, it has a diameter of about 25 km and an albedo from 0.24 to 0.43 (Ali-Lagoa et al., 2018; Masiero et al. 2014, Usui et al. 2011). For the first time photometric observations were carried out by Behrend et al. (2020) in 2005. They obtained the rotation period of $6^h.014$ with the lightcurve amplitude of 0.31 mag. Following observations were performed to determine the rotational properties and the shape of this asteroid (Brinsfield et al., 2010; Hanus et al., 2016; Klinglesmith et al, 2014; etc.).

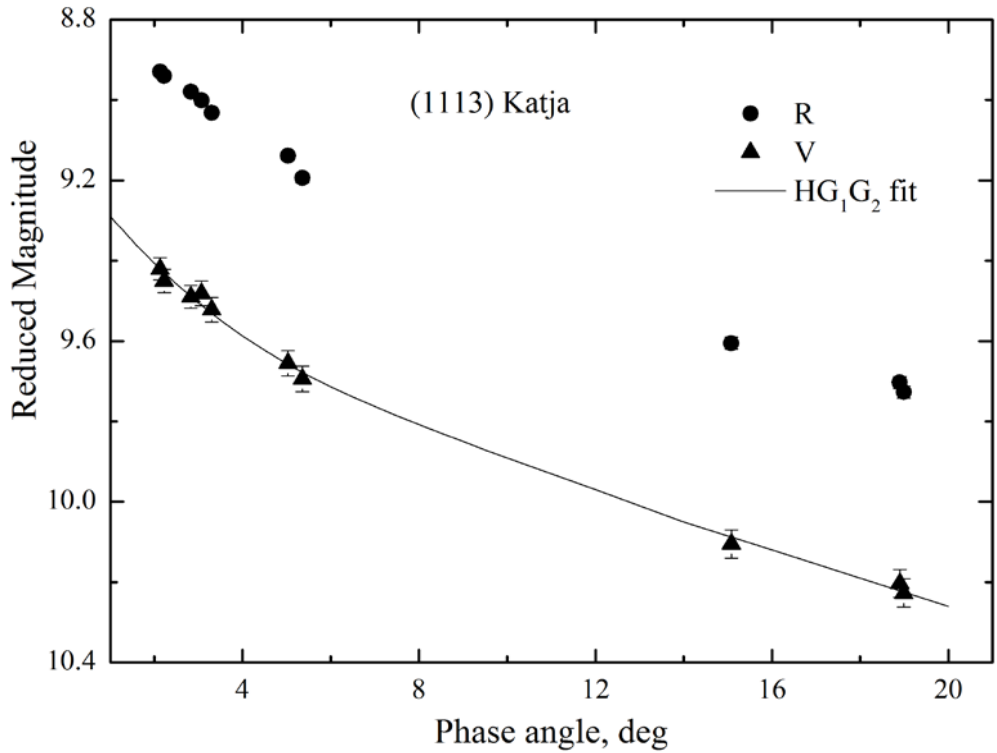

Fig. 20. Magnitude-phase relation of asteroid (1113) Katja (solid line is the best fit with the $HG_1G_2$-function).

We performed observations of the asteroid during eight nights in four months in 2015. Our data allowed us to make a composite lightcurve with a period equal to $6^h.0143 \pm 0.0005$ (Fig. 21). The observations almost completely cover the asteroid rotation. The lightcurve amplitude in that opposition was about 0.50 mag. Our rotation period is close to the previously determined one. We also measured the V-R color index that is 0.46 mag. Our observations allowed to construct the magnitude-phase dependence that is presented in Fig. 22 in the *V* and *R* bands. The solid line indicates the approximation of the phase curve with the $HG_1G_2$-function with the parameters: $H = 10.08$ mag, $G_1 = 0.16$, $G_2 = 0.50$ (see Table 2). It should be noted that our estimate of the absolute magnitude (*H*=10.33) differs from the value given by Veres et al. (2015) (*H*=10.06) and by the MPC (*H*=10.02). The linear phase coefficient $\beta$ is equal to $0.026 \pm 0.002$ mag/deg, and the amplitude of the OE can be estimated as about 0.31 mag.

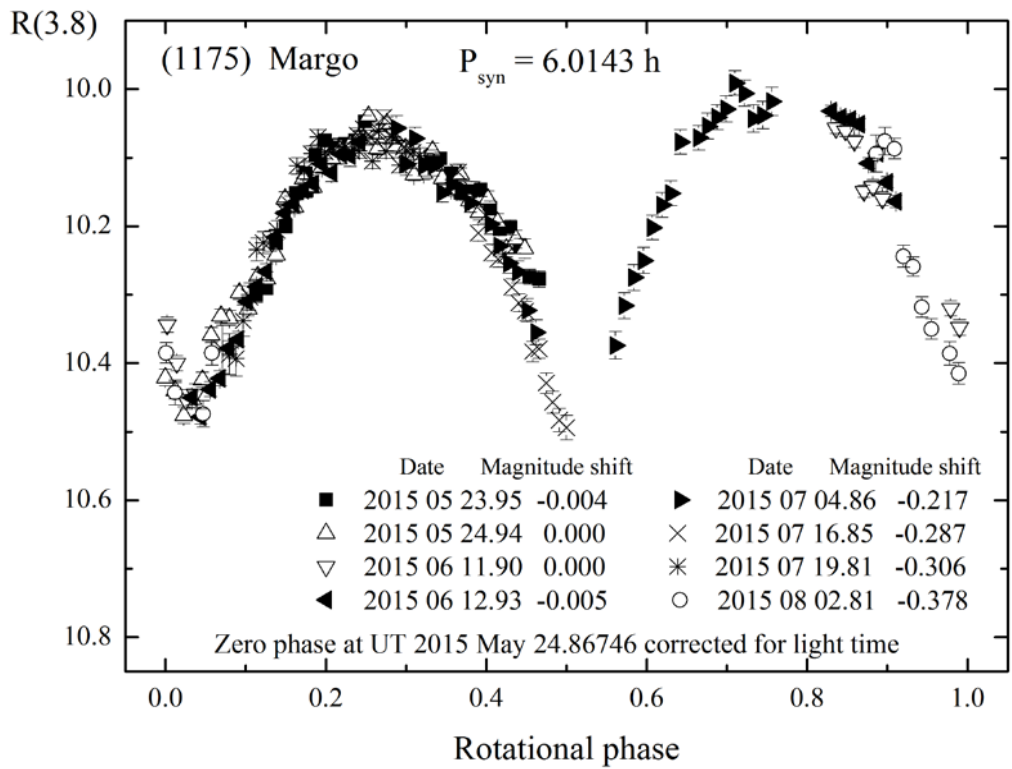

Fig. 21. Composite lightcurve of asteroid (1175) Margo.

*(2057) Rosemary.* This object has a diameter of about 12 km and an albedo ranging from 0.12 to 0.14 (Tedesco et al., 2002; Massiero et al., 2012) Its orbit has a semi-major axis of 3.09 au, an eccentricity 0.23 and small inclination of 1°.4. The only physical parameters for this asteroid are the *H* magnitudes given by MPC (*H*=12.59) and by Veres et al. (2015) (*H*=12.61). We did not find any other published photometric observations of this object. Our observations for nine nights were not enough to determine the rotation period due to the small lightcurve amplitude (<0.05 mag), although sometimes there were variations of about 0.1 mag (Fig. 23). This indicates a complex rotation of the asteroid. Additional observations are needed to investigate the rotational properties of this object in more detail. The magnitude-phase relations in the *V* and *R* bands were constructed using an average brightness during each night (Fig. 24).

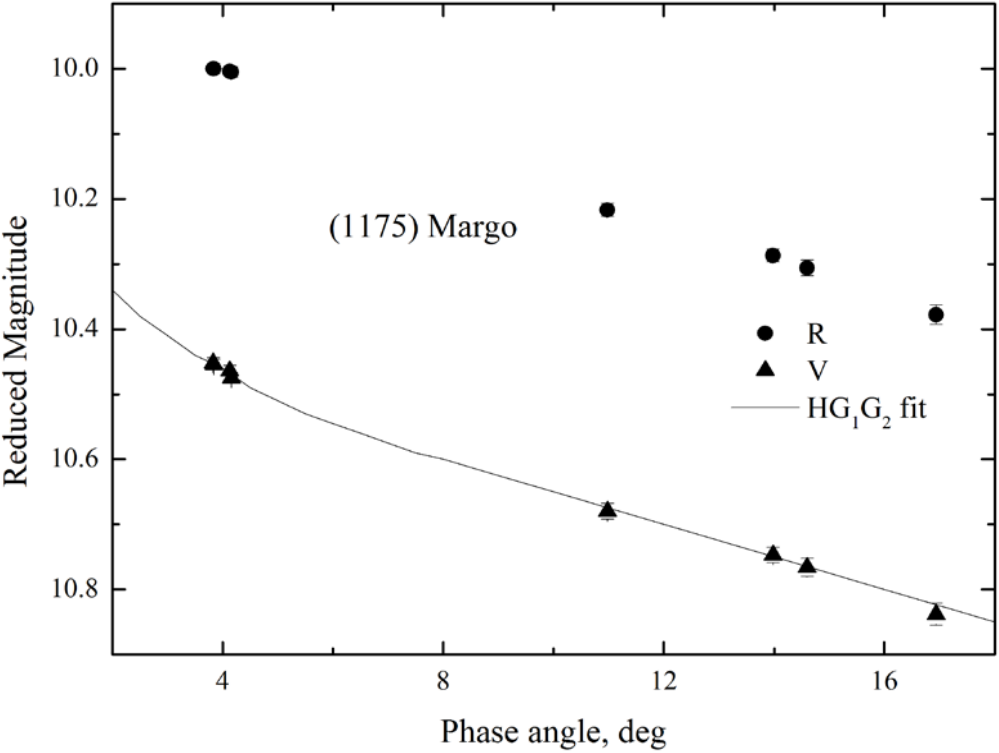

Fig. 22. Magnitude-phase relation of asteroid (1175) Margo (solid line is the best fit with the $HG_1G_2$-function).

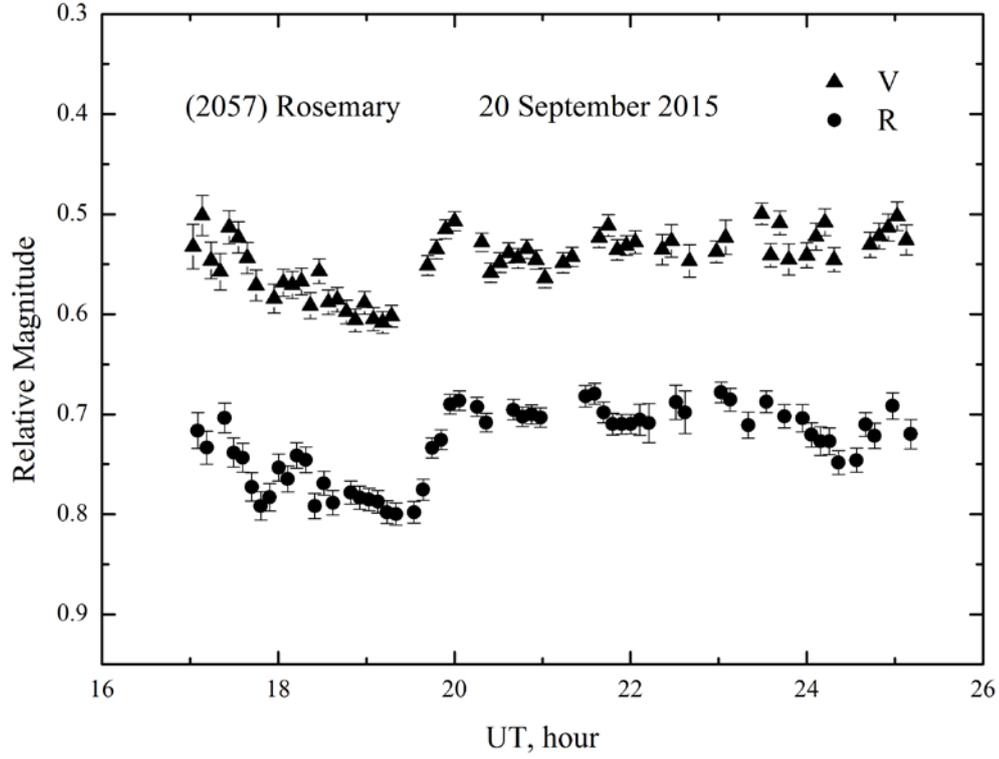

Fig. 23. Lightcurve of asteroid (2057) Rosemary on 20 September 2015.

The solid line indicates the approximation of the phase curve with the $HG_1G_2$-function with the parameters $H = 12.72$ mag, $G_1 = 0.41$, $G_2 = 0.32$ (see Table 2). Our estimate of the absolute magnitude differ from the value given by Veres et al. (2015) ($H=12.61$) and from the MPC value ($H=12.59$). The linear phase coefficient $\beta$ is equal to $0.033 \pm 0.002$ mag/deg, and the OE amplitude can be estimated as about 0.22 mag.

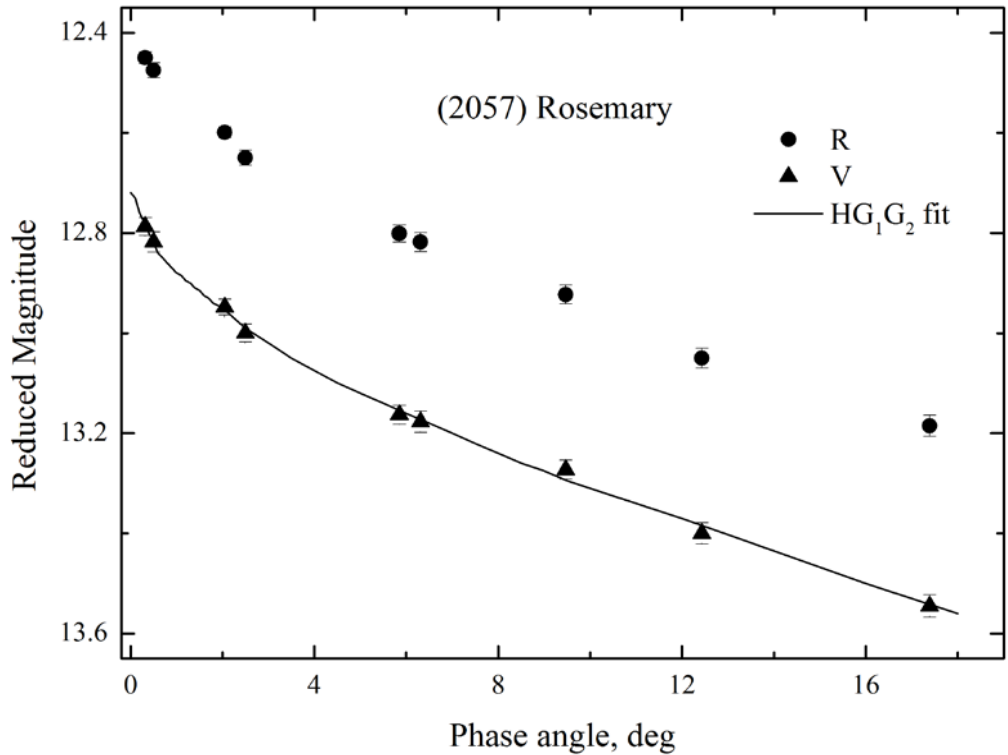

Fig. 24. Magnitude-phase relation of asteroid (2057) Rosemary (solid line is the best fit with the $HG_1G_2$-function).

## 3. Analysis of magnitude-phase curves, spectral data and albedos of the investigated asteroids

We compared our data on magnitude-phase relations of the investigated asteroids with average magnitude-phase relations for high (E-type), moderate (S-complex) and low albedo (C-complex) asteroids according to Shevchenko et al. (2019a). The data were shifted in the magnitude-axis in order to obtain the best fit with the average phase curves. Zero point of relative magnitudes was set to correspond to a phase angle of 8º where the opposition effect typically starts to become noticeable (Belskaya, Shevchenko 2000). It should be noted that a good coverage of the observational data both in the OE region and in the linear part is extremely important for a correct analysis of magnitude-phase curves.

As seen from Fig. 25, the asteroid (260) Huberta has the phase curve as for the average magnitude- phase relation of low albedo asteroids and (1113) Katja shows the phase relation also close to those of low albedo asteroids. Thus, our observations confirmed previously determined albedo and composition type of (260) Huberta. As for (1113) Katja its type was not previously determined. Our data assume lower albedo than published ones while *V-R*=0.49 mag is within the range for S-type asteroids. Further observations of this asteroid are needed to understand a possible peculiarity of its surface.

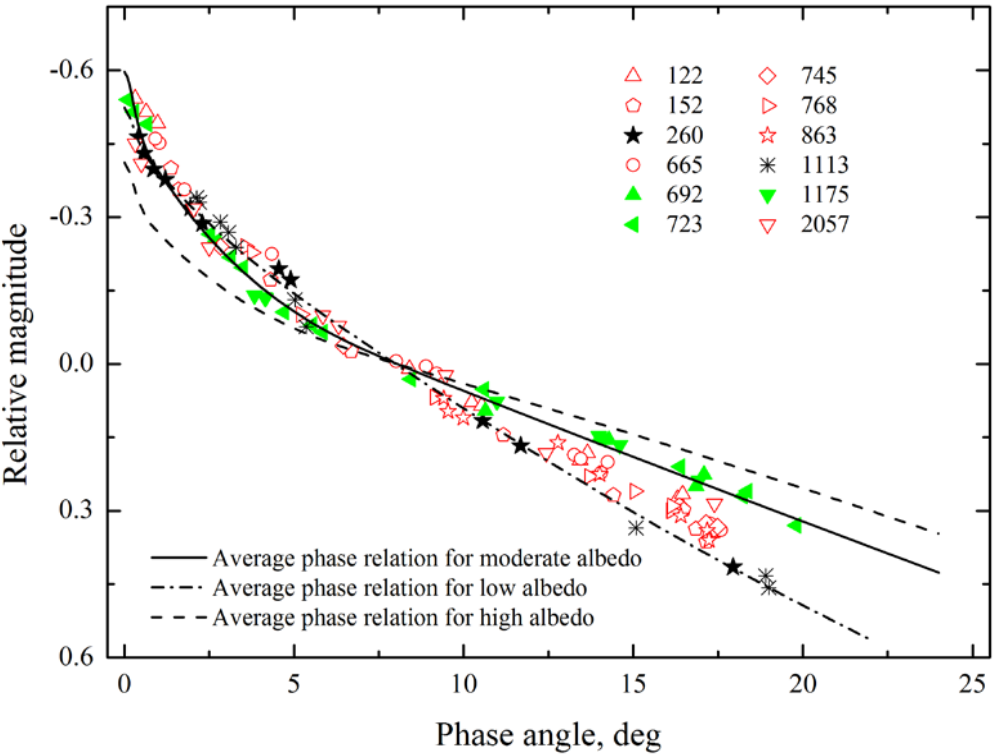

Fig. 25. Comparison of the magnitude-phase relations of the investigated asteroids with average magnitude-phase relations for high, moderate and low albedo asteroids.

Other investigated asteroids have demonstrated the brightness behavior similar to moderate albedo asteroids. Phase curves of three asteroids (692) Hippodamia, (723) Hammonia, and (1175) Margo practically coincide with the average phase curve for the S-complex asteroids, assuming their S-type classification. The available spectra for (692) Hippodamia (Vilas et al. 1992, Bus, Binzel 2002, Lazarro et al. 2004, De Pra et al. 2018) have well detected absorption band of pyroxene-olivine silicates near 0.9-1.0 μm, which are typical for S-type surfaces. As for (723) Hammonia our data contradict the previously determined C-type in Bus and Binzel (2002) but support conclusions made by Kasuga et al. (2013). They modelled the spectral data for (723) Hammonia and concluded about the presence of around 52% amorphous carbon, and 48% amorphous pyroxene (with 70% Mg) with grain size 40 μm suggesting a moderate albedo (Kasuga et al. 2013). In addition, in a recent article by Hasegawa et al. (2021), this asteroid was classified as Xn-type. The classification of (1175) Margo based on SDSS data (Carvano et al. 2010) is confirmed by our data.

Asteroids (122) Gerda, (152) Atala, (665) Sabine, (745) Mauritia, (768) Struveana, and (2057) Rosemary show typical for moderate-albedo asteroids behavior in the OE region but their phase slopes are steeper compared to the average S-type slope (see Fig.25). It may point out that these asteroids have surface properties different from those typical for S-type asteroids. These asteroids may belong to other moderate albedo composition classes.

The spectrum of (122) Gerda (Devogèle et al., 2018) has two small absorption bands in the near infrared that are characteristically for pyroxene-olivine silicates but they are not as pronounced as those of typical S-types. Also, polarimetric properties of (122) Gerda differ from the S-class and are close to those of K-type (Belskaya et al., 2017). Additional observations are needed to classify this asteroid.

The visible spectrum of (152) Atala has a linear spectral slope of 10.6 %/$10^3$ Å and an absorption band around 0.90 μm, typical for pyroxenes-dominant materials (Bus and Binzel 2002). However, its polarimetric properties are different from the S-class and close to A-type, respectively (Belskaya et al., 2017). Moreover, the measured phase behavior of (152) Atala is close to that measured for the A-type asteroid (863) Benkoela (see Fig. 25). Spectral observations of (152) Atala in the infrared range are needed to confirm its possible A-type assumed from photometric and polarimetric data.

Asteroids (665) Sabine and (768) Struveana classified as belonging to X-type show almost identical phase-angle behavior and most probably belong to the same taxonomic class M (in Tholen's definition). The measured *V-R* color for the previously unclassified asteroid (2057) Rosemary indicates possible M-type composition. The visible spectra of (665) Sabine and (768) Struveana have no noticeable features, also supporting the M-type classification (Lazarro et al., 2004; Clark et al., 2004). Clark et al. (2004) also classified those objects as plausible M-types despite a weak absorption band found at 0.9 μm.

The visible spectrum of (863) Benkoela has a linear slope of 23.3 %/1000 Å and two well-detected absorption bands around 0.9-1.0 μm and 2 μm that are associated with olivine-pyroxene silicates. It was classified as an A-type asteroid (Tholen 1989; Bus and Binzel 2002; DeMeo et al., 2009), which typically have a moderate albedo. Our observations confirmed a moderate-albedo and classification different from S-type.

In Table 3, we summarized available albedo estimations of the investigated asteroids obtained from the infrared data of IRAS, WISE and AKARI satellites in the frame of different thermophysical models. One can see that for some asteroids the differences in published albedos are rather large. In Table 3, we also present the results determined magnitude- phase curve parameters (linear phase coefficient and amplitude of OE).

Fig. 26 shows a diagram of the OE amplitudes versus the linear slopes for investigated asteroids in comparison with available high-quality measurements of main asteroid types (Belskaya, Shevchenko 2000). The asteroid (260) Huberta is a typical low albedo asteroid and falls in the corresponding region on diagram and all others are located in a region for moderate albedo asteroids.

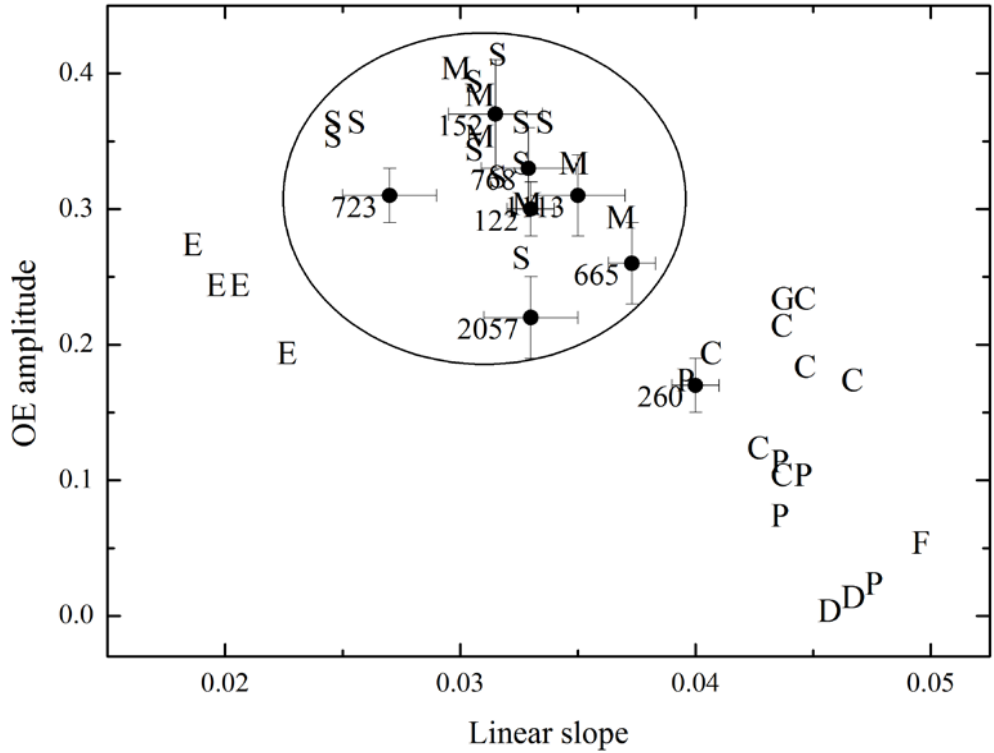

Fig. 26. Diagram of the OE amplitudes versus the linear slopes (data taken from Belskaya, Shevchenko (2000), the investigated asteroids are denoted with filled circles).

Using new albedo sets from WISE and AKARI data (Masiero et al., 2012, 2014, Usui et al., 2011) and new data on linear slopes, we made new calibration of the relation between linear phase coefficients and albedo for the set of asteroids for which high-quality phase curves were obtained. We obtained the following expression:

$\beta = 0.016\ (\pm 0.001) - 0.022\ (\pm 0.001) \log p_v$,

that was previously presented in Belskaya, Shevchenko (2018). Using this expression, we estimated the geometric albedo of investigated asteroids based on the measured values of their linear coefficients. These data are presented in Table 3. The albedo estimations of the investigated asteroids are confirmed their affiliation to moderate albedo asteroids. In Fig. 27, we presents the diagram of albedos obtained in this work in comparison with average values from thermal data listed in Table 3.

In general, albedo estimations from photometric data agree with albedos obtained from infrared data within the uncertainties. There are two noticeable exceptions. Albedos of asteroids (665) Sabine and (863) Benkoela are significantly lower than those obtained from the infrared data. Measured phase-angle curves of these asteroids exclude high-albedo surfaces.

In case of asteroid (692) Hippodamia, our uncertainty of the obtained albedo is high so additional observations are needed to clarify the albedo of this asteroid.

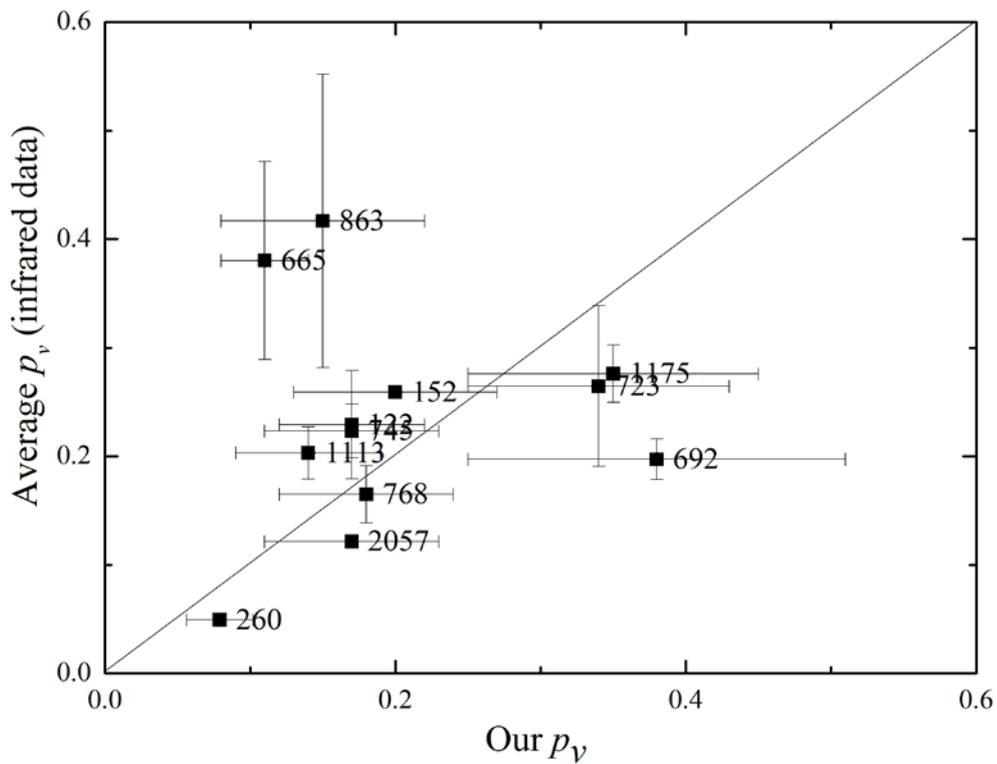

Fig. 27. Comparison of albedos estimated from phase curve and infrared data.

## 4. Conclusions

As a result of the long-term photometric observations we have measured the magnitude-phase relations for twelve asteroids from the outer part of the main belt. Eleven asteroids of our sample have moderate and high albedos according to the infrared data. Our observations were aimed to verify albedo and taxonomic class of these asteroids based on their magnitude-phase curves. For 10 of 12 asteroids our albedo estimations are in agreement with the albedos from infrared data within the uncertainties of measurements. Two asteroids, (665) Sabine and (863) Benkoela, show magnitude-phase dependencies typical for moderate-albedo asteroids which are contradictory to their high-albedos obtained by the infrared surveys.

The measured magnitude-phase relations confirm the previously determined spectral types of (260) Huberta (C-type), (692) Hippodamia (S-type) and (1175) Margo (S-type). Asteroids (665) Sabine and (768) Struveana previously classified as X-type show phase-curve behavior typical for moderate-albedo asteroids and may belong to the M-type. The phase-curve of (723) Hammonia is typical for S-type which contradicts the previously determined C-type (Bus and Binzel 2002), although we do not exclude Xn-type (Hasegawa et al. 2021). We confirmed the moderate-albedo of asteroid (122) Gerda and (152) Atala, but their phase-curves are different from typical for the S-type and may indicate more rare compositional types. For asteroids (745) Mauritia, (1113) Katja and (2057) Rosemary the composition types were not previously assigned. Based on phase-curve behaviours and V-R colors, (2057) Rosemary most probably belong to M-type, while asteroids (745) Mauritia and (1113) Katja belong to S-complex. The obtained phase curve of the A-type asteroid (863) Benkoela did not cover small phase angles in the opposition effect range. Further observations of this and other A-type asteroids are needed to understand typical features of phase-curve behaviours of A-type asteroids in comparison with other types.

Thus, our observations of 12 selected asteroids from the outer part of the main belt have shown that a) two asteroids have considerably overestimated albedo from the infrared data; b) most part of moderate albedo asteroids have surface properties different from typical for the S-type asteroids; c) three of the measured asteroids may belong to rare M-type objects.

We have also determined the lightcurve amplitudes of the observed asteroids and obtained new or improved values of the rotation periods for 10 from 12. New, more accurate values of the absolute magnitudes were determined and it can be used for more accurate estimation of the albedos or diameters of these asteroids.

**Acknowledgments**

This work was partially funded by the National Research Foundation of Ukraine, project N2020.02/0371 "Metallic asteroids: search for parent bodies of iron meteorites, sources of extraterrestrial resources". The results of the work are based partially on observational data obtained with the 1 m telescope in Simeiz of the Center for collective use of INASAN. The work of AVK and IVR has been partially supported by the Program BR05236322 of the Ministry of Education and Science of the RK and the scientific and technical program BR05336383 Aerospace Committee of the Ministry of Defense and Aerospace Industry of the Republic of Kazakhstan. Observations at Abastumani were partly supported by the Shota Rustaveli National Science Foundation, Grant RF-18-1193. This research has made use of data and/or services provided by the International Astronomical Union's Minor Planet Center. We are grateful to the reviewers K. Muinonen and R. Gil-Hutton for their constructive comments that improved our article.

Table 1. Aspect data and measured magnitudes and colors of the observed asteroids. The columns present the date of observation, ecliptic coordinates at epoch 2000.0, the distance from the asteroid to the Sun and to the Earth in astronomical unit, the phase angle, the reduced magnitudes corrected for distances from the Earth and the Sun and corresponding to the primary maxima of the asteroid lightcurves, and their errors.

| Name | UT Date day | $\lambda_{2000}$ deg | $\beta_{2000}$ deg | r au | Δ au | α deg | $V_O(1,\alpha)$ mag | | $R_O(1,\alpha)$ mag | |
|---|---|---|---|---|---|---|---|---|---|---|
| 1 | 2 | 3 | 4 | 5 | 6 | 7 | 8 | 9 | 10 | 11 |
| (122) Gerda | 2003 04 09.79 | 198.961 | 0.847 | 3.140 | 2.138 | 0.32 | 7.647 | 0.011 | - | - |
| | 2003 04 10.92 | 198.738 | 0.855 | 3.140 | 2.139 | 0.65 | 7.675 | 0.012 | - | - |
| | 2003 04 11.86 | 198.554 | 0.862 | 3.141 | 2.139 | 0.98 | 7.698 | 0.013 | - | - |
| | 2003 05 02.86 | 194.851 | 0.974 | 3.148 | 2.217 | 8.40 | 8.200 | 0.015 | - | - |
| | 2003 05 08.82 | 194.072 | 0.994 | 3.150 | 2.260 | 10.24 | 8.268 | 0.016 | - | - |
| | 2003 05 09.82 | 193.958 | 0.996 | 3.150 | 2.268 | 10.53 | 8.275 | 0.016 | - | - |
| | 2003 05 20.89 | 193.029 | 1.022 | 3.154 | 2.370 | 13.43 | 8.387 | 0.018 | - | - |
| | 2003 05 21.86 | 192.978 | 1.024 | 3.155 | 2.380 | 13.65 | 8.372 | 0.018 | - | - |
| | 2003 06 04.86 | 192.809 | 1.037 | 3.160 | 2.541 | 16.30 | 8.465 | 0.020 | - | - |
| | 2003 06 05.85 | 192.836 | 1.038 | 3.160 | 2.553 | 16.45 | 8.456 | 0.020 | - | - |
| (152) Atala | 2017 04 14.80 | 207.850 | 4.036 | 3.196 | 2.196 | 1.59 | 8.364 | 0.020 | 7.846 | 0.015 |
| | 2017 04 15.88 | 207.630 | 3.983 | 3.197 | 2.195 | 1.37 | 8.320 | 0.006 | 7.820 | 0.005 |
| | 2017 04 28.84 | 205.034 | 3.296 | 3.206 | 2.219 | 4.31 | 8.548 | 0.015 | 8.076 | 0.012 |
| | 2017 05 05.88 | 203.763 | 2.901 | 3.211 | 2.252 | 6.68 | 8.694 | 0.007 | 8.211 | 0.006 |
| | 2017 05 20.81 | 201.692 | 2.060 | 3.222 | 2.364 | 11.17 | 8.865 | 0.008 | 8.355 | 0.007 |
| | 2017 06 03.83 | 200.736 | 1.314 | 3.232 | 2.512 | 14.41 | 8.988 | 0.009 | 8.485 | 0.008 |
| | 2017 06 15.82 | 200.744 | 0.734 | 3.240 | 2.662 | 16.35 | 9.010 | 0.018 | 8.500 | 0.015 |
| | 2017 06 16.84 | 200.779 | 0.688 | 3.241 | 2.676 | 16.48 | 9.018 | 0.020 | 8.496 | 0.015 |
| | 2017 06 19.83 | 200.910 | 0.554 | 3.243 | 2.716 | 16.84 | 9.057 | 0.022 | 8.558 | 0.012 |
| | 2017 06 21.85 | 201.024 | 0.466 | 3.244 | 2.744 | 17.05 | 9.083 | 0.025 | 8.559 | 0.018 |
| | 2017 06 25.82 | 201.306 | 0.297 | 3.247 | 2.799 | 17.41 | 9.062 | 0.010 | 8.543 | 0.008 |
| (260) Huberta | 2014 09 13.00 | 353.419 | -1.151 | 3.049 | 2.044 | 1.21 | 9.163 | 0.020 | 8.823 | 0.015 |
| | 2014 09 13.99 | 353.231 | -1.183 | 3.049 | 2.044 | 0.86 | 9.142 | 0.018 | 8.802 | 0.012 |
| | 2014 09 14.95 | 353.050 | -1.214 | 3.050 | 2.044 | 0.57 | 9.110 | 0.016 | 8.780 | 0.011 |
| | 2014 09 15.81 | 352.886 | -1.242 | 3.050 | 2.044 | 0.42 | 9.076 | 0.013 | 8.730 | 0.011 |
| | 2014 09 20.93 | 351.919 | -1.404 | 3.050 | 2.050 | 1.95 | 9.221 | 0.015 | 8.876 | 0.012 |
| | 2014 09 21.87 | 351.743 | -1.433 | 3.050 | 2.052 | 2.30 | 9.253 | 0.013 | 8.923 | 0.012 |
| | 2014 09 27.94 | 350.648 | -1.614 | 3.051 | 2.069 | 4.55 | 9.346 | 0.013 | 8.955 | 0.013 |

| 1 | 2 | 3 | 4 | 5 | 6 | 7 | 8 | 9 | 10 | 11 |
|---|---|---|---|---|---|---|---|---|---|---|
|  | 2014 09 28.89 | 350.483 | -1.641 | 3.052 | 2.073 | 4.90 | 9.368 | 0.015 | 8.988 | 0.012 |
|  | 2014 10 15.83 | 348.096 | -2.071 | 3.055 | 2.178 | 10.56 | 9.656 | 0.013 | 9.262 | 0.013 |
|  | 2014 10 19.75 | 347.728 | -2.154 | 3.056 | 2.212 | 11.68 | 9.707 | 0.013 | 9.304 | 0.013 |
|  | 2014 11 23.78 | 348.209 | -2.663 | 3.066 | 2.629 | 17.94 | 9.955 | 0.013 | 9.537 | 0.013 |
| (665) Sabine | 2018 01 14.86 | 129.600 | -2.499 | 3.445 | 2.487 | 4.35 | 8.885 | 0.032 | 8.485 | 0.013 |
|  | 2018 01 22.87 | 127.988 | -2.936 | 3.435 | 2.455 | 1.77 | 8.754 | 0.022 | 8.354 | 0.013 |
|  | 2018 01 24.83 | 127.580 | -3.041 | 3.432 | 2.450 | 1.23 | 8.700 | 0.030 | 8.300 | 0.013 |
|  | 2018 01 25.77 | 127.381 | -3.091 | 3.431 | 2.448 | 1.04 | 8.659 | 0.031 | 8.259 | 0.013 |
|  | 2018 01 26.75 | 127.176 | -3.142 | 3.430 | 2.446 | 0.92 | 8.650 | 0.032 | 8.250 | 0.014 |
|  | 2018 02 19.75 | 122.466 | -4.259 | 3.398 | 2.497 | 8.01 | 9.104 | 0.033 | 8.704 | 0.015 |
|  | 2018 02 22.70 | 121.999 | -4.372 | 3.394 | 2.515 | 8.89 | 9.115 | 0.032 | 8.715 | 0.015 |
|  | 2018 02 23.79 | 121.835 | -4.413 | 3.392 | 2.522 | 9.21 | 9.130 | 0.041 | 8.730 | 0.021 |
|  | 2018 03 11.90 | 120.093 | -4.923 | 3.370 | 2.656 | 13.26 | 9.296 | 0.041 | 8.896 | 0.021 |
|  | 2018 03 12.81 | 120.035 | -4.947 | 3.369 | 2.665 | 13.45 | 9.304 | 0.041 | 8.904 | 0.022 |
|  | 2018 03 15.84 | 119.873 | -5.025 | 3.365 | 2.697 | 14.05 | 9.332 | 0.042 | 8.932 | 0.021 |
|  | 2018 03 16.91 | 119.828 | -5.051 | 3.363 | 2.708 | 14.25 | 9.311 | 0.042 | 8.911 | 0.023 |
|  | 2018 04 03.92 | 119.992 | -5.418 | 3.337 | 2.919 | 16.73 | 9.402 | 0.042 | 9.002 | 0.023 |
|  | 2018 04 04.90 | 120.049 | -5.435 | 3.336 | 2.931 | 16.82 | 9.413 | 0.042 | 9.013 | 0.023 |
|  | 2018 04 05.80 | 120.106 | -5.450 | 3.334 | 2.942 | 16.90 | 9.433 | 0.042 | 9.033 | 0.023 |
|  | 2018 04 09.82 | 120.406 | -5.514 | 3.328 | 2.993 | 17.20 | 9.442 | 0.042 | 9.042 | 0.023 |
|  | 2018 04 11.80 | 120.582 | -5.544 | 3.325 | 3.019 | 17.32 | 9.447 | 0.043 | 9.047 | 0.023 |
|  | 2018 04 12.80 | 120.678 | -5.559 | 3.324 | 3.032 | 17.38 | 9.537 | 0.043 | 9.137 | 0.023 |
|  | 2018 04 13.88 | 120.787 | -5.575 | 3.322 | 3.046 | 17.43 | 9.450 | 0.042 | 9.050 | 0.023 |
|  | 2018 04 15.84 | 120.998 | -5.604 | 3.319 | 3.071 | 17.51 | 9.450 | 0.042 | 9.050 | 0.023 |
|  | 2018 04 17.76 | 121.220 | -5.630 | 3.317 | 3.096 | 17.58 | 9.451 | 0.043 | 9.051 | 0.023 |
|  | 2018 05 10.81 | 125.026 | -5.918 | 3.281 | 3.389 | 17.32 | 9.436 | 0.041 | 9.036 | 0.021 |
|  | 2018 05 11.81 | 125.233 | -5.929 | 3.280 | 3.402 | 17.27 | 9.426 | 0.041 | 9.026 | 0.021 |
| (692) Hippodamia | 2015 05 24.82 | 208.248 | 15.844 | 3.366 | 2.509 | 10.64 | 9.696 | 0.008 | 9.220 | 0.005 |
|  | 2015 06 12.83 | 206.796 | 13.222 | 3.397 | 2.718 | 14.28 | 9.755 | 0.010 | 9.300 | 0.008 |
|  | 2015 07 08.83 | 207.762 | 9.926 | 3.437 | 3.088 | 16.85 | 9.850 | 0.012 | 9.330 | 0.010 |
|  | 2015 07 17.81 | 208.786 | 8.926 | 3.451 | 3.228 | 17.08 | 9.826 | 0.015 | 9.346 | 0.012 |
| (723) Hammonia | 2014 08 08.94 | 348.293 | 0.714 | 2.936 | 2.027 | 10.60 | 10.492 | 0.020 | 10.142 | 0.012 |
|  | 2014 08 30.90 | 344.547 | 0.175 | 2.924 | 1.920 | 2.52 | 10.176 | 0.012 | 9.825 | 0.011 |

| 1 | 2 | 3 | 4 | 5 | 6 | 7 | 8 | 9 | 10 | 11 |
|---|---|---|---|---|---|---|---|---|---|---|
|  | 2014 09 05.88 | 343.318 | 0.018 | 2.921 | 1.913 | 0.10 | 9.890 | 0.011 | 9.550 | 0.010 |
|  | 2014 09 06.82 | 343.123 | -0.007 | 2.921 | 1.913 | 0.28 | 9.929 | 0.022 | 9.575 | 0.020 |
|  | 2014 09 07.78 | 342.924 | -0.033 | 2.921 | 1.913 | 0.67 | 9.950 | 0.022 | 9.600 | 0.020 |
|  | 2014 09 12.86 | 341.881 | -0.167 | 2.918 | 1.918 | 2.72 | 10.189 | 0.020 | 9.839 | 0.015 |
|  | 2014 09 13.87 | 341.678 | -0.194 | 2.918 | 1.920 | 3.13 | 10.207 | 0.020 | 9.872 | 0.015 |
|  | 2014 09 14.78 | 341.495 | -0.218 | 2.917 | 1.922 | 3.49 | 10.213 | 0.018 | 9.893 | 0.013 |
|  | 2014 09 17.82 | 340.900 | -0.296 | 2.916 | 1.930 | 4.70 | 10.304 | 0.015 | 9.984 | 0.012 |
|  | 2014 09 19.97 | 340.492 | -0.352 | 2.915 | 1.937 | 5.53 | 10.360 | 0.014 | 10.010 | 0.013 |
|  | 2014 09 20.76 | 340.346 | -0.372 | 2.914 | 1.940 | 5.84 | 10.350 | 0.016 | 10.025 | 0.014 |
|  | 2014 09 27.78 | 339.142 | -0.546 | 2.911 | 1.973 | 8.44 | 10.457 | 0.016 | 10.121 | 0.015 |
|  | 2014 10 15.77 | 337.144 | -0.940 | 2.903 | 2.110 | 14.06 | 10.585 | 0.016 | 10.245 | 0.015 |
|  | 2014 10 25.75 | 336.844 | -1.123 | 2.898 | 2.213 | 16.37 | 10.648 | 0.016 | 10.300 | 0.015 |
|  | 2014 11 05.78 | 337.218 | -1.295 | 2.893 | 2.342 | 18.22 | 10.705 | 0.020 | 10.360 | 0.018 |
|  | 2014 11 06.82 | 337.291 | -1.310 | 2.893 | 2.355 | 18.36 | 10.695 | 0.020 | 10.350 | 0.018 |
|  | 2014 11 23.67 | 339.275 | -1.521 | 2.886 | 2.574 | 19.80 | 10.770 | 0.020 | 10.420 | 0.018 |
| (745) Mauritia | 2018 01 14.70 | 106.951 | -5.564 | 3.160 | 2.185 | 2.84 | 10.800 | 0.015 | 10.380 | 0.012 |
|  | 2018 01 25.70 | 104.819 | -4.853 | 3.157 | 2.219 | 6.46 | 11.003 | 0.015 | 10.583 | 0.012 |
|  | 2018 03 11.83 | 101.848 | -1.877 | 3.147 | 2.653 | 17.15 | 11.365 | 0.020 | 10.945 | 0.015 |
|  | 2018 03 14.84 | 102.047 | -1.704 | 3.147 | 2.693 | 17.48 | 11.372 | 0.020 | 10.952 | 0.020 |
|  | 2018 04 03.80 | 104.475 | -0.672 | 3.143 | 2.967 | 18.55 | 11.410 | 0.02 | 10.990 | 0.020 |
| (768) Struveana | 2016 04 09.98 | 191.863 | 9.615 | 3.556 | 2.572 | 3.57 | - | -- | 10.000 | 0.005 |
|  | 2016 04 10.89 | 191.680 | 9.563 | 3.557 | 2.575 | 3.76 | - | - | 10.012 | 0.008 |
|  | 2016 04 16.82 | 190.540 | 9.209 | 3.566 | 2.601 | 5.21 | - | - | 10.139 | 0.011 |
|  | 2016 05 01.95 | 188.106 | 8.193 | 3.586 | 2.710 | 9.16 | - | - | 10.308 | 0.012 |
|  | 2016 05 24.87 | 186.308 | 6.559 | 3.616 | 2.972 | 13.71 | - | - | 10.470 | 0.015 |
|  | 2016 06 04.82 | 186.339 | 5.817 | 3.629 | 3.127 | 15.04 | - | - | 10.500 | 0.015 |
|  | 2016 06 24.81 | 187.757 | 4.594 | 3.652 | 3.433 | 16.11 | - | - | 10.540 | 0.018 |
|  | 2016 06 25.82 | 187.870 | 4.538 | 3.653 | 3.449 | 16.13 | - | - | 10.540 | 0.018 |
|  | 2016 06 26.82 | 187.986 | 4.482 | 3.654 | 3.464 | 16.14 | - | - | 10.530 | 0.020 |
| (863) Benkoela | 2018 10 06.06 | 57.376 | -31.719 | 3.295 | 2.594 | 14.01 | 9.927 | 0.018 | 9.385 | 0.017 |
|  | 2018 10 14.04 | 56.464 | -32.355 | 3.295 | 2.532 | 12.78 | 9.870 | 0.020 | 9.321 | 0.014 |
|  | 2018 11 03.00 | 52.792 | -33.132 | 3.293 | 2.432 | 9.99 | 9.830 | 0.017 | 9.270 | 0.016 |
|  | 2018 11 08.90 | 51.464 | -33.075 | 3.293 | 2.421 | 9.54 | 9.823 | 0.017 | 9.257 | 0.015 |

| 1 | 2 | 3 | 4 | 5 | 6 | 7 | 8 | 9 | 10 | 11 |
|---|---|---|---|---|---|---|---|---|---|---|
|  | 2018 11 12.85 | 50.553 | -32.955 | 3.293 | 2.418 | 9.41 | 9.772 | 0.037 | 9.230 | 0.014 |
|  | 2019 01 12.80 | 43.527 | -25.216 | 3.287 | 2.827 | 16.40 | 10.053 | 0.032 | 9.470 | 0.014 |
|  | 2019 01 24.73 | 44.559 | -23.327 | 3.286 | 2.976 | 17.18 | 10.078 | 0.027 | 9.500 | 0.027 |
|  | 2019 01 25.73 | 44.679 | -23.172 | 3.286 | 2.989 | 17.23 | 10.091 | 0.023 | 9.520 | 0.013 |
| (1113) Katja | 2016 08 30.00 | 345.125 | 6.099 | 3.132 | 2.133 | 3.30 | 9.522 | 0.023 | 9.032 | 0.010 |
|  | 2016 09 03.90 | 344.118 | 6.397 | 3.125 | 2.122 | 2.22 | 9.450 | 0.029 | 8.940 | 0.009 |
|  | 2016 09 04.84 | 343.923 | 6.452 | 3.124 | 2.120 | 2.13 | 9.420 | 0.028 | 8.930 | 0.008 |
|  | 2016 09 10.81 | 342.677 | 6.783 | 3.116 | 2.117 | 2.83 | 9.490 | 0.028 | 8.980 | 0.008 |
|  | 2016 09 11.71 | 342.491 | 6.830 | 3.114 | 2.117 | 3.07 | 9.481 | 0.031 | 9.001 | 0.011 |
|  | 2016 09 17.98 | 341.223 | 7.133 | 3.106 | 2.126 | 5.03 | - | - | 9.139 | 0.011 |
|  | 2016 09 18.96 | 341.030 | 7.176 | 3.104 | 2.129 | 5.36 | 9.694 | 0.032 | 9.194 | 0.012 |
|  | 2016 10 22.68 | 336.839 | 8.055 | 3.057 | 2.354 | 15.08 | 10.105 | 0.035 | 9.605 | 0.015 |
|  | 2016 11 21.76 | 338.409 | 8.157 | 3.015 | 2.703 | 18.90 | 10.203 | 0.035 | 9.703 | 0.015 |
|  | 2016 11 23.68 | 338.675 | 8.155 | 3.012 | 2.728 | 18.99 | 10.227 | 0.035 | 9.727 | 0.015 |
| (1175) Margo | 2015 05 23.95 | 254.213 | 5.569 | 3.198 | 2.203 | 4.13 | 10.464 | 0.008 | 10.004 | 0.005 |
|  | 2015 05 24.94 | 254.024 | 5.644 | 3.197 | 2.200 | 3.83 | 10.452 | 0.008 | 10.000 | 0.005 |
|  | 2015 06 11.90 | 250.430 | 6.894 | 3.185 | 2.186 | 3.84 | 10.455 | 0.008 | 10.000 | 0.008 |
|  | 2015 06 12.93 | 250.228 | 6.957 | 3.185 | 2.188 | 4.16 | 10.475 | 0.008 | 10.005 | 0.008 |
|  | 2015 07 04.86 | 246.659 | 8.015 | 3.171 | 2.295 | 10.98 | 10.680 | 0.012 | 10.217 | 0.011 |
|  | 2015 07 16.85 | 245.598 | 8.377 | 3.163 | 2.399 | 13.98 | 10.747 | 0.012 | 10.287 | 0.011 |
|  | 2015 07 19.81 | 245.454 | 8.446 | 3.161 | 2.428 | 14.60 | 10.766 | 0.014 | 10.306 | 0.012 |
|  | 2015 08 02.81 | 245.425 | 8.688 | 3.152 | 2.584 | 16.94 | 10.838 | 0.017 | 10.378 | 0.015 |
| (2057) Rosemary | 2015 09 19.97 | 357.433 | -0.767 | 2.446 | 1.442 | 0.50 | 12.833 | 0.021 | 12.491 | 0.015 |
|  | 2015 09 20.88 | 357.247 | -0.757 | 2.445 | 1.441 | 0.32 | 12.787 | 0.018 | 12.450 | 0.012 |
|  | 2015 09 24.95 | 356.418 | -0.709 | 2.440 | 1.439 | 2.05 | 12.929 | 0.016 | 12.582 | 0.011 |
|  | 2015 09 25.89 | 356.228 | -0.698 | 2.438 | 1.439 | 2.50 | 13.010 | 0.018 | 12.662 | 0.015 |
|  | 2015 10 02.92 | 354.859 | -0.610 | 2.429 | 1.447 | 5.85 | 13.162 | 0.019 | 12.801 | 0.017 |
|  | 2015 10 03.91 | 354.677 | -0.598 | 2.428 | 1.449 | 6.31 | 13.179 | 0.021 | 12.822 | 0.019 |
|  | 2015 10 10.84 | 353.507 | -0.507 | 2.419 | 1.471 | 9.47 | 13.270 | 0.019 | 12.923 | 0.019 |
|  | 2015 10 17.86 | 352.570 | -0.414 | 2.411 | 1.504 | 12.43 | 13.430 | 0.021 | 13.082 | 0.021 |
|  | 2015 10 31.84 | 351.667 | -0.235 | 2.397 | 1.600 | 17.39 | 13.540 | 0.022 | 13.185 | 0.021 |

Table 2. Physical parameters of observed asteroids

| Asteroid | $a$, au | Spec. class | $p_v$ [e] | $D$ [e] km | $P$, hours | Amp., mag | $B-V$, mag | $V-R$, mag | $H$, mag | | $G_1$ | | $G_2$ | |
|---|---|---|---|---|---|---|---|---|---|---|---|---|---|---|
| (122) Gerda | 3.225 | ST[a], L[b] | 0.173 | 85.4 | 10.6958 ± 0.0005 | 0.10 | 0.88 ± 0.02 | 0.45 ± 0.02 | 7.549 | +0.005 / -0.015 | 0.46 | +0.07 / -0.08 | 0.22 | +0.04 / -0.05 |
| (152) Atala | 3.138 | | 0.257 | 57.1 | 6.2447 ± 0.0002 | 0.30 | 0.88 ± 0.03 | 0.49 ± 0.02 | 8.055 | +0.016 / -0.024 | 0.30 | +0.04 / -0.03 | 0.29 | +0.02 / -0.02 |
| (260) Huberta | 3.454 | CX[a], X[g] | 0.054 | 95.2 | 8.2895 ± 0.0002 | 0.28 | - | 0.34 ± 0.02 | 9.021 | +0.018 / -0.018 | 0.70 | +0.03 / -0.03 | 0.10 | +0.07 / -0.07 |
| (665) Sabine | 3.143 | X[g] | 0.365 | 53.0 | 4.29407 ± 0.00005 | 0.34 | 0.69 ± 0.02 | 0.40 ± 0.02 | 8.502 | +0.046 / -0.043 | 0.52 | +0.11 / -0.11 | 0.19 | +0.05 / -0.06 |
| (692) Hippodamia | 3.386 | S[a], Sl[g] | 0.185 | 45.3 | 8.9972 ± 0.0005 | 0.28 | - | 0.48 ± 0.02 | 9.230 | +0.170 / -0.120 | 0.15 | +0.02 / -0.02 | 0.60 | +0.05 / -0.05 |
| (723) Hammonia | 2.993 | C[b] | 0.294 | 28.3 | 5.4349 ± 0.0002 | 0.10 | - | 0.35 ± 0.02 | 9.876 | +0.012 / -0.012 | 0.24 | +0.03 / -0.03 | 0.43 | +0.02 / -0.02 |
| (745) Mauritia | 3.260 | | 0.249 | 23.2 | 9.9425 ± 0.0005 | 0.40 | - | 0.42 ± 0.02 | 10.430 | +0.090 / -0.080 | 0.32 | +0.12 / -0.11 | 0.31 | +0.04 / -0.04 |
| (768) Struveana | 3.138 | X[a], X[g] | 0.169 | 31.2 | 10.7458 ± 0.0005 | 0.20 | - | - | 9.409 | +0.061 / -0.057 | 0.19 | +0.08 / -0.07 | 0.29 | +0.02 / -0.03 |
| (863) Benkoela | 3.203 | A[bf] | 0.444 | 31.5 | - | 0.05 | 1.06 ± 0.04 | 0.56 ± 0.03 | 9.160 | +0.57 / -0.57 | 0.26 | +0.04 / -0.04 | 0.37 | +0.05 / -0.05 |
| (1113) Katja | 3.110 | | 0.211 | 38.2 | 18.465 ± 0.005 | 0.10 | - | 0.49 ± 0.02 | 9.050 | +0.075 / -0.079 | 0.36 | +0.14 / -0.12 | 0.20 | +0.06 / -0.06 |
| (1175) Margo | 3.214 | S[h] | 0.302 | 23.0 | 6.01425 ± 0.0005 | 0.25 | - | 0.46 ± 0.02 | 10.080 | +0.190 / -0.173 | 0.16 | +0.26 / -0.14 | 0.50 | +0.04 / -0.04 |
| (2057) Rosemary | 3.087 | - | 0.124[d] | 12.5[d] | - | 0.10 | - | 0.36 ± 0.02 | 12.720 | +0.050 / -0.053 | 0.41 | +0.04 / -0.05 | 0.32 | +0.05 / -0.03 |

a) Tholen, (1989); b) Bus and Binzel, (2002); c) Ali-Lagoa et al., (2018); d) Masiero et al., (2012); e) Usui et al., (2011); f) DeMeo et al. (2009); g) Lazzaro et al. (2004); h) Carvano et al. (2010)

Table 3. Results of albedo determinations of observed asteroids

| Asteroid | $p_v$ Tedesco et al., (2002) | $p_v$ Masiero et al., (2012) | $p_v$ Masiero et al., (2014) | $p_v$ Usui et al., (2011) | $p_v$ Ali-Lagoa et al., (2018) | $\beta$ mag/deg | OE mag | $p_v$ this work |
|---|---|---|---|---|---|---|---|---|
| (122) Gerda | 0.188 ± 0.009 | 0.297 ± 0.032 | 0.251 ± 0.037 | 0.173 ± 0.006 | 0.236 ± 0.047 | 0.033 ± 0.001 | 0.30 ± 0.02 | 0.17 ± 0.05 |
| (152) Atala | - | 0.257 ± 0.018 | 0.237 ± 0.044 | 0.257 ± 0.010 | 0.263 ± 0.053 | 0.032 ± 0.002 | 0.37 ± 0.04 | 0.20 ± 0.07 |
| (260) Huberta | 0.051 ± 0.004 | - | 0.044 ± 0.010 | 0.054 ± 0.002 | 0.048 ± 0.011 | 0.040 ± 0.001 | 0.17 ± 0.02 | 0.079 ± 0.023 |
| (665) Sabine | 0.390 ± 0.039 | 0.494 ± 0.042 | - | 0.365 ± 0.012 | 0.272 ± 0.052 | 0.037 ± 0.001 | 0.26 ± 0.03 | 0.11 ± 0.03 |
| (692) Hippodamia | 0.179 ± 0.015 | - | 0.205 ± 0.029 | 0.185 ± 0.006 | 0.220 ± 0.040 | 0.025 ± 0.003 | 0.21 ± 0.05 | 0.38 ± 0.13 |
| (723) Hammonia | 0.183 ± 0.015 | 0.352 ± 0.048 | - | 0.294 ± 0.031 | 0.229 ± 0.092 | 0.027 ± 0.002 | 0.31 ± 0.02 | 0.34 ± 0.09 |
| (745) Mauritia | - | - | 0.200 ± 0.023 | 0.249 ± 0.032 | 0.221 ± 0.088 | 0.032 ± 0.002 | 0.28 ± 0.03 | 0.17 ± 0.06 |
| (768) Struveana | - | - | 0.137 ± 0.028 | 0.169 ± 0.012 | 0.189 ± 0.060 | 0.033 ± 0.002 | 0.33 ± 0.03 | 0.18 ± 0.06 |
| (863) Benkoela | 0.595 ± 0.070 | 0.790 ± 0.333 | 0.290 ± 0.032 | 0.444 ± 0.027 | 0.338 ± 0.135 | 0.034 ± 0.004 | 0.26 ± 0.03 | 0.15 ± 0.07 |
| (1113) Katja | 0.207 ± 0.023 | 0.168 ± 0.026 | 0.195 ± 0.018 | 0.211 ± 0.008 | 0.234 ± 0.047 | 0.035 ± 0.002 | 0.31 ± 0.03 | 0.14 ± 0.05 |
| (1175) Margo | - | - | 0.249 ± 0.042 | 0.302 ± 0.026 | 0.277 ± 0.111 | 0.026 ± 0.002 | 0.31 ± 0.03 | 0.35 ± 0.10 |
| (2057) Rosemary | 0.119 ± 0.018 | 0.124 ± 0.027 | - | - | - | 0.033 ± 0.002 | 0.22 ± 0.03 | 0.17 ± 0.06 |